\documentclass[twocolumn, showpacs,amsmath,amssymb,aps,pra,floatfix,longbibliography, 10pt]{revtex4-1}

\usepackage[latin1]{inputenc}
\usepackage[american]{babel}
\usepackage[T1]{fontenc}
\usepackage{graphicx}
\usepackage{amsfonts}
\usepackage{amsmath}
\usepackage{mathrsfs}
\usepackage{hyperref}
\usepackage{epstopdf}
\usepackage{bm, epsfig}

\usepackage{mathtools}
\usepackage{afterpage}

\DeclarePairedDelimiterX\braket[2]{\langle}{\rangle}{#1 \delimsize\vert #2}
\hypersetup{colorlinks=true, urlcolor=blue, citecolor=blue, filecolor=blue, linkcolor=blue}
\newcommand*{\diff}{\mathop{}\!\mathrm{d}}

\newcommand*{\Imm}{\mathop{}\!\mathbf{Im}}

\newcommand{\uimm}{\mathrm{i}}
\newcommand{\eu}{\mathrm{e}}

\begin{document}

\title{Transient-absorption phases with strong probe and pump pulses}

\author{Vadim~Becquet}
\altaffiliation[On leave from: ]{Aix Marseille Univ, Centrale Marseille, Marseille, France}
\affiliation{Max-Planck-Institut f\"ur Kernphysik, Saupfercheckweg 1, 69117 Heidelberg, Germany}
\author{Stefano~M.~Cavaletto}
\email[Email: ]{smcavaletto@gmail.com}
\affiliation{Max-Planck-Institut f\"ur Kernphysik, Saupfercheckweg 1, 69117 Heidelberg, Germany}
\date{\today}
\begin{abstract}
The quantum dynamics of a system of Rb atoms, modeled by a $V$-type three-level system interacting with intense probe and pump pulses, are studied. The time-delay-dependent transient-absorption spectrum of an intense probe pulse is thus predicted, when this is preceded or followed by a strong pump pulse. Numerical results are interpreted in terms of an analytical model, which allows us to quantify the oscillating features of the resulting transient-absorption spectra in terms of the atomic populations and phases generated by the intense pulses. Strong-field-induced phases and their influence on the resulting transient-absorption spectra are thereby investigated for different values of pump and probe intensities and frequencies, focusing on the atomic properties which are encoded in the absorption line shapes for positive and negative time delays.
\end{abstract}


\maketitle

\section{Introduction}

Phases represent the essential feature of any wave-like phenomena, lying at the heart of coherence and interference effects in classical and quantum physics. In atoms and molecules, phases define the shape of a wave packet in a superposition of quantum states and hence determine its subsequent time evolution. Manipulating atomic and molecular dynamics with external electromagnetic fields \cite{1367-2630-12-7-075008, tannor2007introduction, doi:10.1146/annurev.physchem.040808.090427, doi:10.1146/annurev.physchem.59.032607.093818}, e.g., by using strong femto- or attosecond pulses \cite{PhysRevLett.86.47, PhysRevLett.94.083002, PhysRevLett.100.233603, PhysRevLett.102.023004, PhysRevA.81.063410}, requires full control of the generated quantum phases. However, traditional spectroscopy methods usually do not provide access to the phase information: for instance, for nonautoionizing bound states, absorption spectra typically consist of Lorentzian lines, with spectral intensities quantifying the atomic populations. 

The manipulation of absorption line shapes in transient-absorption-spectroscopy experiments \cite{Mathies06051988, doi:10.1146/annurev.pc.43.100192.002433, PhysRevLett.98.143601, GoulielmakisNature466, PhysRevLett.106.123601} has been recently identified as a key mechanism to gain access to atomic and molecular phase dynamics. Absorption lines originate from the interference between a probe pulse transmitting through the medium and the field emitted by the system \cite{RevModPhys.40.441}. The dipole response of the system and, consequently, the resulting absorption spectrum can be modified by applying an intense pump pulse, preceding or following the probe pulse at variable time delays \cite{0953-4075-49-6-062003, PhysRevLett.105.143002, PhysRevA.86.063408, Chini-SciRep, Ott10052013, 1367-2630-16-11-113016, PhysRevLett.112.103001, Meyer22122015, PhysRevA.94.023403, stooss_reconstructing}. Thereby, symmetric Lorentzian absorption lines are converted into Fano-like lines, with time-delay-dependent features quantifying the population and phase modification induced by the interaction with the strong pump pulse. 

When interpreting spectral line-shape changes in terms of the underlying atomic dynamics, the action of the weak probe pulse is usually assumed as a small, well understood perturbation. The attention is thus focused on the characterization of the action of the pump pulse as a function of its parameters such as, e.g., intensity and laser frequency, and the main line-shape modifications are exclusively attributed to its nonlinear interaction with the system. Recent investigations of transient-absorption spectra in Rb atoms were based on this assumption \cite{PhysRevLett.115.033003, Liu-SciRep}. 

Here, in contrast, we fully account for the effect of a potentially intense probe pulse, investigating how the population and phase changes induced by both pulses are encoded in its absorption spectrum. On the one hand, this allows us to fully interpret transient-absorption spectra in terms of the pump and probe parameters of interest, without \textit{a priori} assumptions, which may not correspond to the conditions featured in an experiment and, hence, could lead to an inappropriate or incomplete reconstruction of the strong-field dynamics of the system. On the other hand, by considering cases in which pump and probe pulses exhibit the same intensities, we can highlight the essential differences between spectra where the probe, i.e., measured, pulse either precedes or follows the pump pulse. A proper interpretation of transient-absorption spectra is crucial for the extraction of strong-field dynamical information from these spectra, and the implementation of recently suggested deterministic strong-field quantum-control methods \cite{cavaletto2016deterministic}.

We use a $V$-type three-level scheme to model an ensemble of Rb atoms, with the $5s\,^2S_{1/2}\rightarrow 5p\,^2P_{1/2}$ ($794.76\,\mathrm{nm}$) and $5s\,^2S_{1/2}\rightarrow 5p\,^2P_{3/2}$ ($780.03\,\mathrm{nm}$) transitions excited by femtosecond pump and probe pulses of variable intensities and time delays. In Sec.~\ref{Theoretical model}, we present the theoretical model used to describe the evolution of the system and to predict the associated transient-absorption spectra. The numerical results are presented in Sec.~\ref{Results and discussion}. In particular, time-delay-dependent transient-absorption spectra are shown in Subsec.~\ref{Transient-absorption spectra for intense pump and probe pulses} for different pump- and probe-pulse intensities. An analytical model based on recently introduced interaction operators \cite{cavaletto2016deterministic} is used in Subsec.~\ref{Interpretation of pump- and probe-pulse-induced phases in terms of interaction-operator matrix elements} to interpret the numerical results, focusing on the atomic-phase information which can be extracted from the spectra for different intensities and laser frequencies of the pump and probe pulses. Section~\ref{Conclusion}
summarizes the results obtained. Atomic units are used throughout unless otherwise stated.

\section{Theoretical model}

\label{Theoretical model}

\subsection{Three-level model and equations of motion}
\label{Three-level model and equations of motion}

\begin{figure}[t]
\includegraphics[width= \columnwidth]{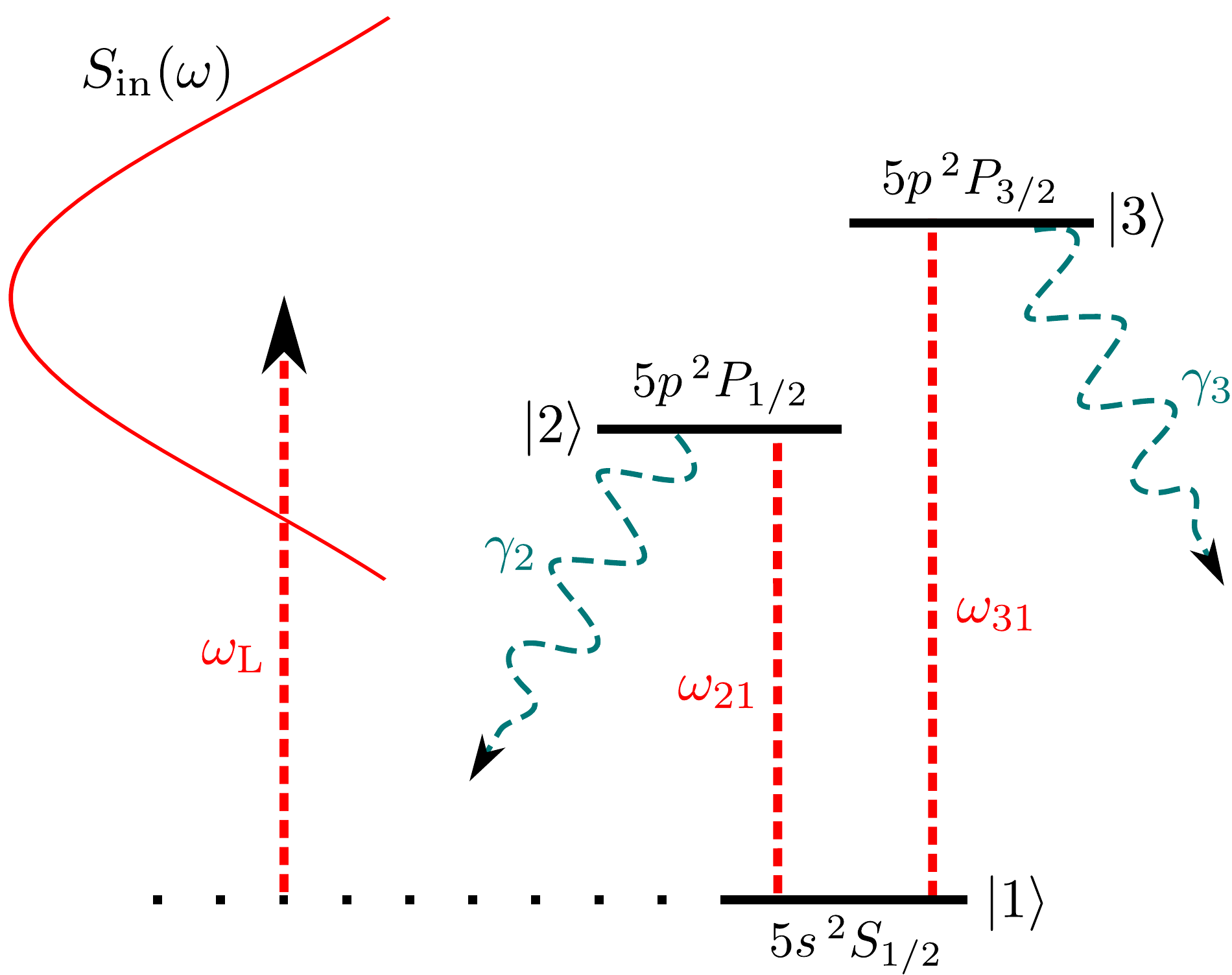}
\caption{$V$-type three-level scheme, with transitions energies $\omega_{21} = 1.56\,\mathrm{eV}$ and $\omega_{31} = 1.59\,\mathrm{eV}$ and decay rates $\gamma_{2} = \gamma_3 = 1/(500\,\mathrm{fs})$, used to model Rb atoms interacting with broadband laser pulses of frequency $\omega_{\mathrm{L}}$ and spectral intensity $S_{\mathrm{in}}(\omega)$.}
\label{fig:threelevelsystem}
\end{figure}

\begin{figure}[t]
\includegraphics[width= 0.9\columnwidth]{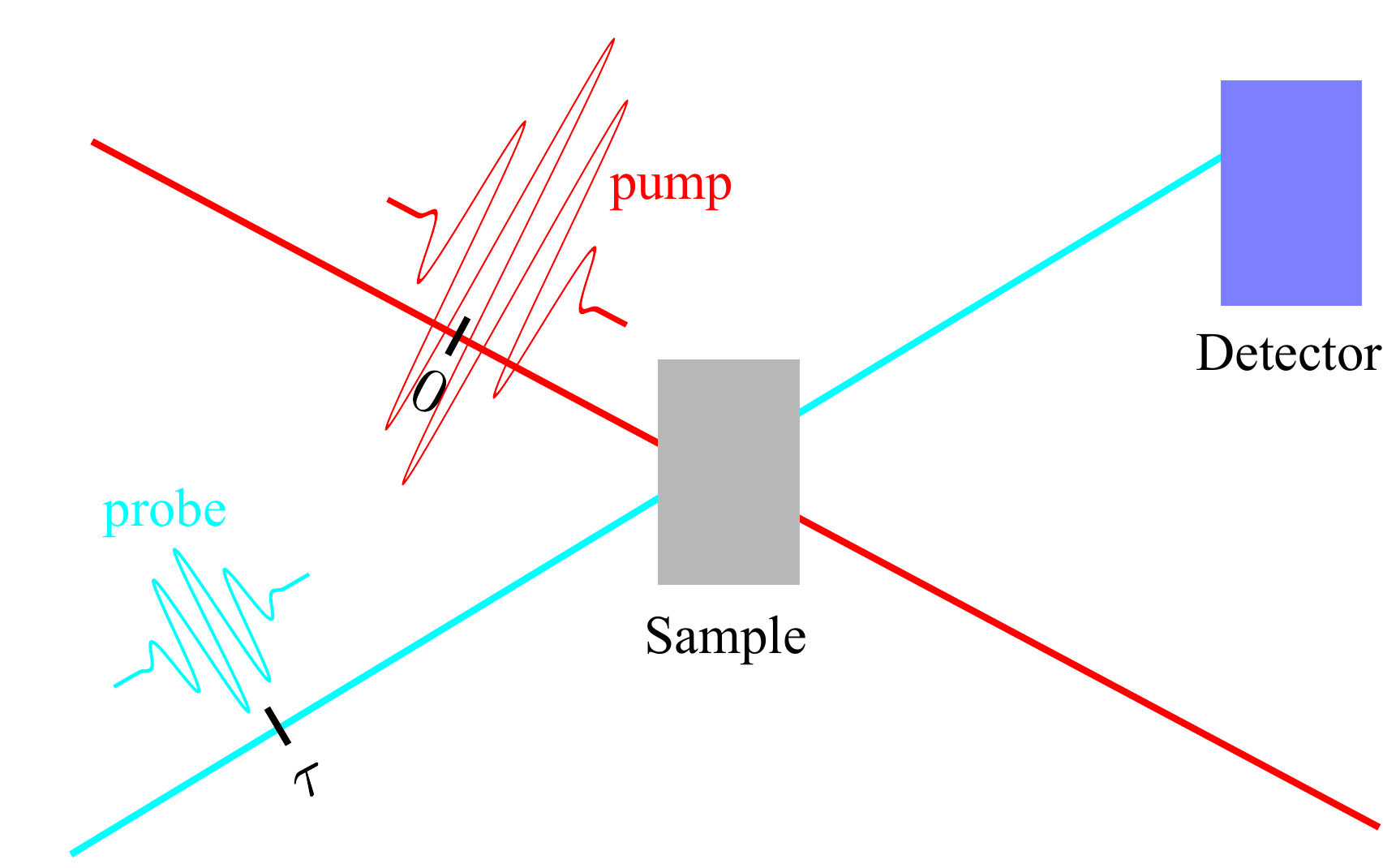}
\caption{Experimental setup for the detection of the optical-density transient-absorption spectrum of a transmitted probe pulse, delayed by $\tau$ with respect to a pump pulse, in a noncollinear geometry.}
\label{fig:experiment}
\end{figure}

We consider the $V$-type three-level system depicted in Fig.~\ref{fig:threelevelsystem}, modeling fine-structure-split $5s\,^2S_{1/2}\rightarrow 5p\,^2P_{1/2}$ and $5s\,^2S_{1/2}\rightarrow 5p\,^2P_{3/2}$ transitions in Rb atoms \cite{PhysRevA.30.2881, PhysRevA.65.043406, PhysRevA.69.022509}. In particular, we introduce the state 
\begin{equation}
|\psi(t,\tau)\rangle = \sum_{i=1}^3 c_i(t,\tau)|i\rangle,
\end{equation}
written in terms of the ground state $|1\rangle \equiv 5s\,^2S_{1/2}$ and the excited states $|2\rangle \equiv 5p\,^2P_{1/2}$ and $|3\rangle \equiv 5p\,^2P_{3/2}$, with associated quantum amplitudes $c_i(t,\tau)$ and energies $\omega_i$, $i\in\{1,\,2,\,3\}$. The system interacts with a pump pulse, centered on $t = 0$ and modeled by the classical field
\begin{equation}
\boldsymbol{\mathcal{E}}_{\mathrm{pu}}(t) = \mathcal{E}_{\mathrm{pu},0}\,f(t)\,\cos(\omega_{\mathrm{L}}t)\hat{\boldsymbol{e}}_z, 
\end{equation}
and a delayed probe pulse, centered on time delay $t = \tau$ and similarly described as
\begin{equation}
\boldsymbol{\mathcal{E}}_{\mathrm{pr}}(t) = \mathcal{E}_{\mathrm{pr},0}\,f(t-\tau)\,\cos[\omega_{\mathrm{L}}(t-\tau)]\hat{\boldsymbol{e}}_z\,,
\end{equation}
as shown in Fig.~\ref{fig:experiment}.
Both pulses are aligned along the polarization vector $\hat{\boldsymbol{e}}_z$, have the same frequency $\omega_{\mathrm{L}}$, vanishing carrier-envelope phases, and intensities $I_{\mathrm{pu/pr}} = \mathcal{E}_{\mathrm{pu/pr},0}^2/(8\pi\alpha)$ related to the peak field strengths $\mathcal{E}_{\mathrm{pu/pr},0}$ via the fine-structure constant $\alpha$. We model their envelope functions as
\begin{equation}
f(t) = \left\{
\begin{aligned}
&\cos^2(\pi t/T) &\text{if } &|t|\leq T/2,\\
&0 &\text{if } &|t|> T/2,
\end{aligned}
\right.
\end{equation}
with $T = \pi T_{\mathrm{FWHM}}/(2\arccos\sqrt[4]{1/2})$ and $T_{\mathrm{FWHM}} = 30\,\mathrm{fs}$, defined as the full width at half maximum (FWHM) of $|f(t)|^2$ \cite{diels2006ultrashort}. Positive time delays correspond to a typical pump-probe setup, in which the system is first excited by the pump pulse and the resulting dynamics are measured by a probe pulse. In contrast, negative time delays describe experiments in which the dipole response generated by the first arriving probe pulse is subsequently modified by the pump pulse, resulting in an intensity- and time-delay-dependent modulation of the line shape of the absorption spectrum of the transmitted probe pulse. 

The linearly polarized pulses excite electric-dipole-($E1$-)allowed transitions $|1\rangle\rightarrow |k\rangle$, $k\in\{2,\,3\}$, with equal magnetic quantum number, $\Delta M = 0$, and dipole-moment matrix elements $\boldsymbol{D}_{1k} = D_{1k}\hat{\boldsymbol{e}}_z$. The formulas are written for general complex values of $D_{1k}$, although these are real and positive for our atomic implementation with Rb atoms, with $D_{12} = 1.75\,\mathrm{a.u.}$ and $D_{13} = 2.47\,\mathrm{a.u.}$ \cite{PhysRevA.65.043406}. For the intensities considered here, we neglect the presence of higher excited states, to which states $|2\rangle$ and $|3\rangle$ could also be coupled. The total Hamiltonian of the system
\begin{equation}
\hat{H} = \hat{H}_0 + \hat{H}_{\mathrm{int}}(t,\tau)
\end{equation}
then consists of the unperturbed atomic Hamiltonian
\begin{equation}
\hat{H}_0 = \sum_{i = 1}^3 (\omega_i-\uimm\gamma_i/2)|i\rangle\langle i|
\label{eq:H0}
\end{equation}
and the $E1$ light-matter interaction Hamiltonian in the rotating-wave approximation \cite{Scully:QuantumOptics, Kiffner_review, Foot:AtomicPhysics}
\begin{equation}
\hat{H}_{\mathrm{int}} = -\frac{1}{2}\sum_{k = 2}^3 \varOmega_{\mathrm{R}k}(t,\tau)\,\eu^{\uimm \omega_{\mathrm{L}}t}\,|1\rangle\langle k| + \mathrm{H.c.}
\label{eq:Hint}
\end{equation}
In Eq.~(\ref{eq:H0}), the complex eigenvalues $(\omega_i-\uimm\gamma_i/2)$ of $\hat{H}_0$ are given by the energies $\omega_i$ and the decay rates $\gamma_i$, included in order to effectively account for broadening effects in the experiment and defining an effective time scale for the dipole decay \cite{Ott10052013, 0953-4075-49-6-062003}. Transition energies $\omega_{ij} = \omega_i - \omega_j$ are equal to $\omega_{21} = 1.56\,\mathrm{eV}$ and $\omega_{31} = 1.59\,\mathrm{eV}$ \cite{PhysRevA.30.2881, PhysRevA.65.043406, PhysRevA.69.022509}, whereas we set $\gamma_1 = 0$ and $\gamma_2 = \gamma_3 = 1/(500\,\mathrm{fs})$. In Eq.~(\ref{eq:Hint}), the time- and time-delay-dependent Rabi frequencies have been introduced \cite{Scully:QuantumOptics}:
\begin{equation}
\varOmega_{\mathrm{R}k}(t,\tau) = D_{1k}[\mathcal{E}_{\mathrm{pr},0}f(t-\tau)\eu^{-\uimm \omega_{\mathrm{L}}\tau} + \mathcal{E}_{\mathrm{pu},0}f(t)].
\end{equation}
The equations of motion (EOMs) satisfied by the vector
\begin{equation}
\vec{c}(t,\tau) = (c_1(t,\tau),\,c_2(t,\tau),\,c_3(t,\tau))^{\mathrm{T}},
\end{equation}
of components given by the amplitudes of the state vector $|\psi(t,\tau)\rangle$, are determined by the Schr\"odinger equation
\begin{equation}
\uimm \frac{\diff |\psi(t,\tau)\rangle}{\diff t} = \hat{H} |\psi(t,\tau)\rangle,
\end{equation}
which leads to
\begin{equation}
\frac{\mathrm{d}\vec{c}}{\mathrm{d}t} = 
\begin{pmatrix}
   0 & \uimm\frac{\varOmega_{\mathrm{R}2}}{2}\,\eu^{\uimm\omega_{\mathrm{L}} t} & \uimm\frac{\varOmega_{\mathrm{R}3}}{2}\,\eu^{\uimm\omega_{\mathrm{L}} t} \\
   \uimm\frac{\varOmega_{\mathrm{R}2}^*}{2}\,\eu^{-\uimm\omega_{\mathrm{L}} t} & -\frac{\gamma_2}{2}-\uimm\omega_{21} & 0 \\
   \uimm\frac{\varOmega_{\mathrm{R}3}^*}{2}\,\eu^{-\uimm\omega_{\mathrm{L}} t} & 0 & -\frac{\gamma_3}{2}-\uimm\omega_{31}
\end{pmatrix}
\vec{c}\,.
\label{eq:matrixC}
\end{equation}
The system is assumed to be initially in its ground state $|\psi_0\rangle = |1\rangle$, i.e., $c_{i,0} = \delta_{i1}$.

\subsection{Transient-absorption spectrum}
We solve the EOMs in Eq.~(\ref{eq:matrixC}) in order to simulate experimental optical-density (OD) absorption spectra
\begin{equation}
\mathcal{S}_{\mathrm{exp}}(\omega, \tau) = -\log\left[\frac{S_{\mathrm{pr,out}}(\omega,\tau)}{S_{\mathrm{pr,in}}(\omega)}\right],
\end{equation}
where $S_{\mathrm{pr,in}}(\omega,\tau)$ is the spectral intensity of the incoming probe pulse, whereas $S_{\mathrm{pr,out}}(\omega,\tau)$ is that of the transmitted probe pulse, explicitly dependent upon the time delay between pump and probe pulses. For low densities and small medium lengths, where propagation effects can be neglected, the time-delay-dependent absorption spectrum $\mathcal{S}_{\mathrm{exp}}(\omega,\tau)$ can be calculated in terms of the single-particle dipole response of the system \cite{RevModPhys.40.441}
\begin{equation}
\begin{aligned}
&\mathcal{S}_1(\omega) \propto \\
&-\omega\Imm\left[\frac{\sum_{k=2}^3 D_{1k}^*\int_{-\infty}^\infty c_1(t,\tau) \,c_k^*(t,\tau)\,\eu^{-\uimm\omega t}\,\diff t}{\int_{-\infty}^{\infty}\mathcal{E}^-_{\mathrm{pr}}(t)\,\eu^{-\uimm \omega t} \diff t}\right],
\end{aligned}
\label{eq:spectruminitial}
\end{equation}
where
\begin{equation}
\mathcal{E}^-_{\mathrm{pr}}(t) = \frac{1}{2}\mathcal{E}_{\mathrm{pr},0}\,f(t-\tau)\,\eu^{\uimm\omega_{\mathrm{L}}(t-\tau)}
\end{equation}
is the negative-frequency complex electric field \cite{diels2006ultrashort} and $c_1(t,\tau) \,c_k^*(t,\tau)$ here represents the dipole response of the $k$th transition. In the following calculations, the denominator in Eq.~(\ref{eq:spectrum1}) is approximated by 
\begin{equation}
\begin{aligned}
&\int_{-\infty}^{\infty}\mathcal{E}^-_{\mathrm{pr}}(t)\,\eu^{-\uimm \omega t} \diff t \\
=\,& \eu^{-\uimm\omega\tau}\,\frac{\mathcal{E}_{\mathrm{pr},0}}{2}\,\int_{-\infty}^{\infty} f(t-\tau)\,\eu^{-\uimm(\omega - \omega_{\mathrm{L}})(t-\tau)} \diff t \\
\approx \,& \eu^{-\uimm\omega\tau}\,\frac{\mathcal{E}_{\mathrm{pr},0}}{2} \int_{-\infty}^{\infty} f(t)\, \diff t = \eu^{-\uimm\omega\tau}\,K_{\mathrm{pr}},
\end{aligned}
\end{equation}
which is valid for an incoming probe pulse much broader than the transition energy between the two excited states, such that its spectral intensity can be approximately considered constant in the frequency range of interest. Spectra associated with different probe-pulse intensities, therefore, need to be properly normalized via the multiplication factor $K_{\mathrm{pr}}$ for comparison. Equation~(\ref{eq:spectrum1}) can then be rewritten as
\begin{equation}
\begin{aligned}
&\mathcal{S}_1(\omega) \propto \\
&-\omega\Imm\left[\frac{\sum_{k=2}^3 D_{1k}^*\int_{-\infty}^\infty c_1(t,\tau)\,c_k^*(t,\tau)\,\eu^{-\uimm\omega (t-\tau)}\,\diff t}{K_{\mathrm{pr}}}\right],
\end{aligned}
\label{eq:spectrum1}
\end{equation}
with the Fourier transform in the numerator centered around the arrival time of the probe pulse.

For the noncollinear geometry depicted in Fig.~\ref{fig:experiment}, fast oscillations of the measured transient-absorption spectrum as a function of time delay $\tau$ cannot be distinguished and are averaged out \cite{PhysRevLett.115.033003, Liu-SciRep}. Here, this is taken into account by convolving $S_{1}(\omega,\tau)$ with a normalized Gaussian function $G(\tau,\Delta \tau)$ of width $\Delta \tau = 5\times 2\pi/\omega_L$, which leads to
\begin{equation}
\mathcal{S}(\omega,\tau) = \int_{-\infty}^{\infty}G(\tau - \tau',\Delta\tau)\,\mathcal{S}_1(\omega, \tau')\,\diff \tau'\,.
\label{eq:spectrumaveraged}
\end{equation}

\subsection{Analytical model in terms of interaction operators}
\label{Analytical model in terms of interaction operators}
In order to interpret numerical results from the simulation of $\mathcal{S}(\omega,\tau)$, we employ the recently introduced strong-field interaction operators $\hat{U}(I)$ to model the effect of a pulse of intensity $I$ on the atomic system \cite{cavaletto2016deterministic}. 

\begin{figure}[t]
\includegraphics[width= \columnwidth]{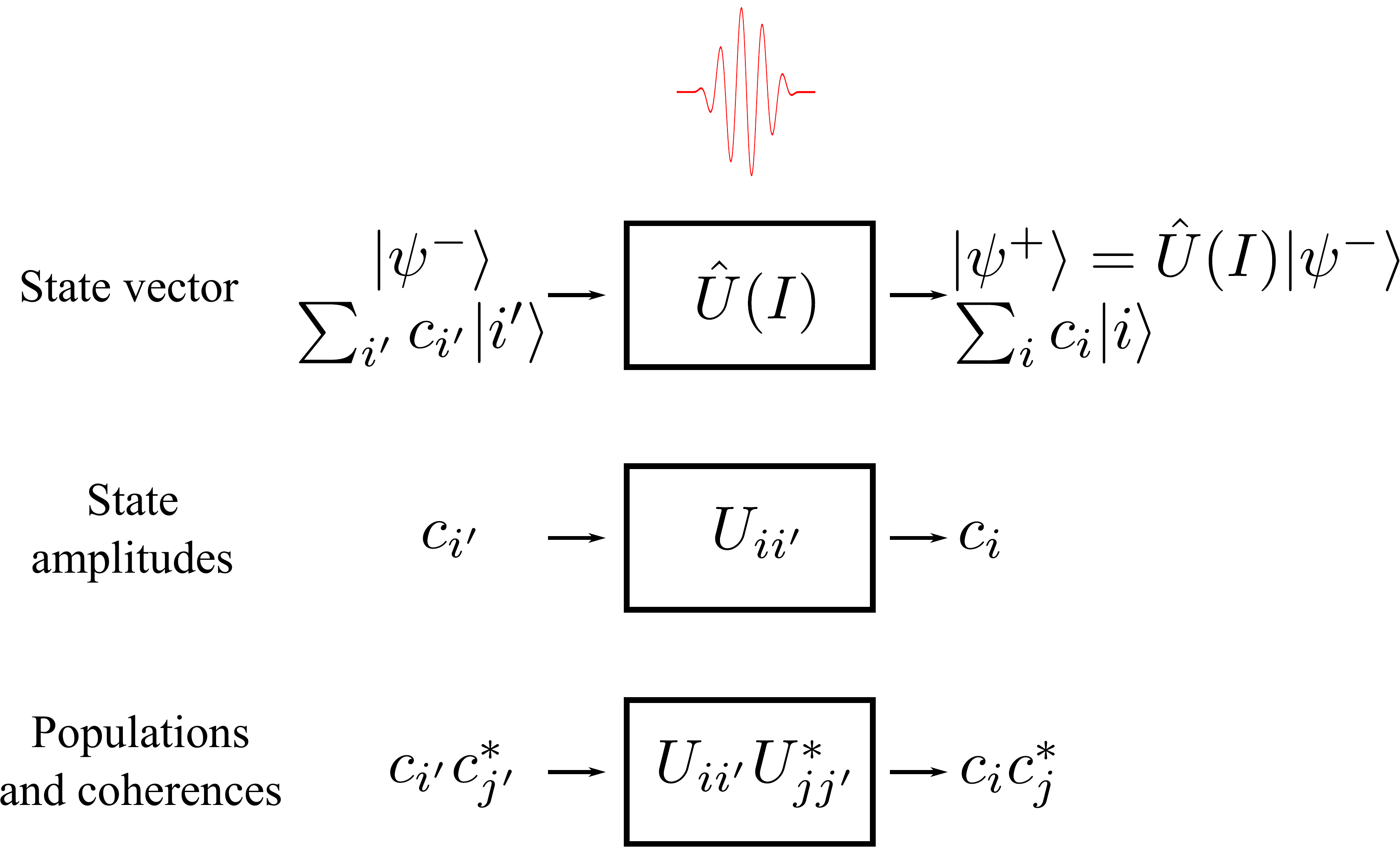}
\caption{Schematic illustration of the interaction operators $\hat{U}(I)$, used to describe the action of an intense pulse on the state of the system, its amplitudes $c_i$, populations $|c_i|^2$, and coherences $c_i c_j^*$, in terms of an effectively instantaneous interaction. The product of matrix elements $U_{ii'}U^*_{jj'}$ describes how the action of the pulse connects the initial population/coherence $c_{i'} c_{j'}^*$ to the final one $c_{i} c_{j}^*$.}
\label{fig:schematic-explanation}
\end{figure}

The time evolution of the system $|\psi(t)\rangle$ from an initial time $t_0$, given by the solution of the EOMs~(\ref{eq:matrixC}), can be written in terms of the evolution operator $\hat{\mathcal{U}}(t,t_0)$,
\begin{equation}
|\psi(t)\rangle = \hat{\mathcal{U}}(t,t_0) |\psi(t_0)\rangle.
\label{eq:Ucal}
\end{equation}
In the absence of external fields, this reduces to the free-evolution operator
\begin{equation}
\hat{V}(t) = \eu^{-\uimm\hat{H}_0 t},
\end{equation}
which describes the dynamics of the unperturbed atomic system. The evolution of the system in the presence of a single pulse of intensity $I = \mathcal{E}_0^2/(8\pi\alpha)$, peak field strength $\mathcal{E}_0$, centered around $t_{\mathrm{c}} = 0$ and with the same envelope $f(t)$ and pulse duration $T$ we introduced in Sec.~\ref{Three-level model and equations of motion}, is then associated with the evolution operator $\hat{\mathcal{U}}_0(t,t_0)$, solution of
\begin{equation}
\begin{aligned}
&\frac{\mathrm{d}\hat{\mathcal{U}}_0(t,t_0)}{\mathrm{d}t}=\\
&\begin{pmatrix}
   0 & \uimm\frac{\varOmega_{\mathrm{R}2}}{2}\,\eu^{\uimm\omega_{\mathrm{L}} t} & \uimm\frac{\varOmega_{\mathrm{R}3}}{2}\,\eu^{\uimm\omega_{\mathrm{L}} t} \\
   \uimm\frac{\varOmega_{\mathrm{R}2}^*}{2}\,\eu^{-\uimm\omega_{\mathrm{L}} t} & -\frac{\gamma_2}{2}-\uimm\omega_{21} & 0 \\
   \uimm\frac{\varOmega_{\mathrm{R}3}^*}{2}\,\eu^{-\uimm\omega_{\mathrm{L}} t} & 0 & -\frac{\gamma_3}{2}-\uimm\omega_{31}
\end{pmatrix}
\hat{\mathcal{U}}_0(t,t_0)\,,
\end{aligned}
\label{eq:matrixU0}
\end{equation}
with initial conditions $\hat{\mathcal{U}}_0(t_0,t_0) = \hat{I}$ and the identity matrix $\hat{I}$. In Eq.~(\ref{eq:matrixU0}), the single-pulse Rabi frequencies $\varOmega_{\mathrm{R}k}(t) = D_{1k}\,\mathcal{E}_0\,f(t)$ are used. For the scheme discussed in this paper, where pump and probe pulses of equal femtosecond duration are employed, the time information related to the continuous evolution of the system in the presence of the pulse can be difficultly extracted. For our purposes, it is therefore beneficial to focus on the \emph{total} action of the pulse, i.e., on the state reached by the system at the conclusion of the interaction with a pulse. Equation~(\ref{eq:matrixU0}) can be used to calculate $\hat{\mathcal{U}}_0(T/2,-T/2)$ and thus connect the initial state $|\psi(-T/2)\rangle$ with the final state $|\psi(T/2)\rangle$ at the end of the pulse:
\begin{equation}
|\psi(T/2)\rangle = \hat{\mathcal{U}}_0(T/2,-T/2) |\psi(-T/2)\rangle.
\end{equation}
However, one can also introduce effective initial ($|\psi^-\rangle$) and final ($|\psi^+\rangle$) states
\begin{equation}
|\psi^{\pm}\rangle = \hat{V}(\mp T/2)|\psi(\pm T/2)\rangle= \eu^{\pm\uimm\hat{H}_0  T/2}|\psi(\pm T/2)\rangle
\end{equation}
and thus define the unique, intensity-dependent interaction operators
\begin{equation}
\hat{U}(I) = \hat{V}(-T/2)\,\hat{\mathcal{U}}_0(T/2,-T/2)\,\hat{V}(-T/2)
\label{eq:definitionU}
\end{equation}
connecting them,
\begin{equation}
|\psi^+\rangle = \hat{U}(I)|\psi^-\rangle,
\end{equation}
thus capturing the essential features of the action of the pulse in terms of an effectively instantaneous interaction, as schematically represented in Fig.~\ref{fig:schematic-explanation}. An analytical model can then be derived to describe the associated $\mathcal{S}(\omega,\tau)$, which enables one to quantify how pulse-induced changes in the population and phase of the atomic states are encoded in observable time-delay-dependent spectra.

For a weak and ultrashort pulse of peak field strength $\mathcal{E}_0$ and envelope $f(t)$, we can introduce approximated Rabi frequencies
\begin{equation}
\varOmega_{\mathrm{R}k}(t)\approx \vartheta_k\,\delta(t),
\end{equation}
with the Dirac $\delta$ and the pulse areas
\begin{equation}
\vartheta_k = \int_{-\infty}^{\infty} D_{1k}\,\mathcal{E}_0\,f(t)\,\diff t.
\end{equation}
The solution of Eq.~(\ref{eq:matrixU0}) and the use of the definition~(\ref{eq:definitionU}) allow one to calculate the associated interaction operator which, up to second order, reads
\begin{equation}
\hat{U}_{\mathrm{weak}} =
\begin{pmatrix}
1-\frac{|\vartheta_2|^2 + |\vartheta_3|^2}{8} & \uimm\frac{\vartheta_2}{2} &  \uimm\frac{\vartheta_3}{2} \\
\uimm\frac{\vartheta_2^*}{2} &1-\frac{|\vartheta_2|^2}{8} & - \vartheta_2^*\vartheta_3\\
\uimm\frac{\vartheta_3^*}{2} &- \vartheta_2\vartheta_3^* &1-\frac{|\vartheta_3|^2}{8} 
\end{pmatrix}.
\label{eq:weak}
\end{equation}

In the following, we interpret intensity-dependent transient-absorption spectra in terms of the matrix elements of pump- and probe-pulse interaction operators for a probe-pump and pump-probe setup. In contrast to previous results \cite{PhysRevLett.115.033003, cavaletto2016deterministic}, population and phase changes due to the interaction with intense probe and pump pulses are both explicitly addressed. Since we are interested in atomic phases, and in particular in their connection with the phase of the time-delay-dependent oscillations displayed by transient-absorption spectra for positive and negative time delays, we do not focus on the case of overlapping pulses. We are therefore allowed to develop an analytical model in which the dynamics of the system are described in terms of well defined sequences of free evolution and interaction with a pump or a probe pulse of given intensity.

\subsubsection{Probe-pump setup}

In a probe-pump setup ($\tau<0$), for nonoverlapping pulses and neglecting the details of the continuous atomic dynamics in the presence of a pulse, the time evolution of the system can be written in terms of the state
\begin{equation}
\begin{aligned}
&|\psi(t,\tau)\rangle = \\
&\left\{
\begin{aligned}
&|\psi_0\rangle, & t<\tau, \\
&\hat{V}(t-\tau)\hat{U}_{\mathrm{pr}}(I_{\mathrm{pr}})|\psi_0\rangle, & \tau<t<0,\\ 
&\hat{V}(t)\hat{U}_{\mathrm{pu}}(I_{\mathrm{pu}})\hat{V}(-\tau)\hat{U}_{\mathrm{pr}}(I_{\mathrm{pr}})|\psi_0\rangle, & t>0,
\end{aligned}
\right.
\end{aligned}
\end{equation}
with $|\psi_0\rangle=|1\rangle$ and where we have introduced the pump- and probe-pulse interaction operators, $\hat{U}_{\mathrm{pu}}(I_{\mathrm{pu}})$ and $\hat{U}_{\mathrm{pr}}(I_{\mathrm{pr}})$, dependent upon the respective pulse intensities. This can be included into Eq.~(\ref{eq:spectrum1}) in order to model the probe-pump spectrum $\mathcal{S}_{1}(\omega,\tau)$, $\tau<0$, in terms of interaction-operator matrix elements. This results in a sum of terms, each of which oscillates as a function of $\tau$ at a given frequency. Thereby, one can recognize, for the frequencies $\omega\approx \omega_{k1}$ in which we are interested, those terms responsible for fast oscillations of $\mathcal{S}_{1}(\omega,\tau)$ as a function of time delay which would not be exhibited by a spectrum measured in a noncollinear geometry. After neglecting these fast oscillating terms, the time-delay-average probe-pump spectrum reads
\begin{equation}
\begin{split}
\mathcal{S}_{\mathrm{prpu}}(\omega,\tau) 
& \propto-\frac{\omega}{K_{\mathrm{pr}}}\Imm\biggl\{\sum_{k=2}^3 \frac{D_{1k}^*}{\uimm(\omega - \omega_{k1}) + \frac{\gamma_k}{2}}\\
&\ \times \bigl[U_{\mathrm{pr},11}U_{\mathrm{pr},k1}^*(1-\eu^{\uimm (\omega - \omega_{k1})\tau}\eu^{\frac{\gamma_k}{2}\tau})\\
&\ \ + U_{\mathrm{pu},11}U_{\mathrm{pu},k2}^* U_{\mathrm{pr},11} U_{\mathrm{pr},21}^* \eu^{\uimm (\omega - \omega_{21})\tau} \eu^{\frac{\gamma_2}{2} \tau}\\
& \ \ + U_{\mathrm{pu},11}U_{\mathrm{pu},k3}^* U_{\mathrm{pr},11} U_{\mathrm{pr},31}^* \eu^{\uimm (\omega - \omega_{31})\tau} \eu^{\frac{\gamma_3}{2} \tau}\bigr]\biggr\}.
\end{split}
\label{eq:Sexpprpu}
\end{equation}

\subsubsection{Pump-probe setup}

When a pump-probe setup is utilized ($\tau>0$), for nonoverlapping pulses and neglecting the details of the continuous atomic dynamics in the presence of a pulse, the atomic state can be modeled as
\begin{equation}
\begin{aligned}
&|\psi(t,\tau)\rangle = \\
&\left\{
\begin{aligned}
&|\psi_0\rangle, & t<0, \\
&\hat{V}(t)\hat{U}_{\mathrm{pu}}(I_{\mathrm{pu}})|\psi_0\rangle, & 0<t<\tau,\\ 
&\hat{V}(t-\tau)\hat{U}_{\mathrm{pr}}(I_{\mathrm{pr}})\hat{V}(\tau)\hat{U}_{\mathrm{pu}}(I_{\mathrm{pu}})|\psi_0\rangle, & t>\tau,
\end{aligned}
\right.
\end{aligned}
\end{equation}
with $|\psi_0\rangle=|1\rangle$. By neglecting fast oscillating terms appearing in the resulting single-particle absorption spectrum~(\ref{eq:spectrum1}) at frequencies $\omega\approx \omega_{k1}$, the time-delay-average pump-probe spectrum can be written in terms of the matrix elements of the interaction operators $\hat{U}_{\mathrm{pu}}(I_{\mathrm{pu}})$ and $\hat{U}_{\mathrm{pr}}(I_{\mathrm{pr}})$ as
\begin{equation}
\begin{split}
&\mathcal{S}_{\mathrm{pupr}}(\omega,\tau) \propto -\frac{\omega}{K_{\mathrm{pr}}}\Imm\biggl[\sum_{k=2}^3\frac{D_{1k}^*}{\uimm(\omega - \omega_{k1}) + \frac{\gamma_k}{2}}\\
&\ \ \ \ \ \ \times \bigl( U_{\mathrm{pr}, 11}U_{\mathrm{pr},k1}^*|U_{\mathrm{pu,}11}|^2 \\
&\ \ \ \ \ \ \ \ \  +U_{\mathrm{pr}, 12}U_{\mathrm{pr},k2}^*|U_{\mathrm{pu,}21}|^2 \,\eu^{-\gamma_2\tau} \\
&\ \ \ \ \ \ \ \ \  +U_{\mathrm{pr}, 12}U_{\mathrm{pr},k3}^*U_{\mathrm{pu,}21} \,U^*_{\mathrm{pu,}31} \,\eu^{\uimm\omega_{32}\tau}\,\eu^{-\frac{\gamma_2+\gamma_3}{2}\tau} \\
&\ \ \ \ \ \ \ \ \  +U_{\mathrm{pr}, 13}U_{\mathrm{pr},k2}^* U_{\mathrm{pu,}31} \,U^*_{\mathrm{pu,}21} \,\eu^{-\uimm\omega_{32}\tau}\,\eu^{-\frac{\gamma_2+\gamma_3}{2}\tau} \\
&\ \ \ \ \ \ \ \ \  +U_{\mathrm{pr}, 13}U_{\mathrm{pr},k3}^*|U_{\mathrm{pu,}31}|^2 \,\eu^{-\gamma_3\tau} \bigr)\biggr].
\end{split}
\label{eq:Sexppupr}
\end{equation}

\section{Results and discussion}
\label{Results and discussion}

\subsection{Transient-absorption spectra for intense probe and pump pulses}
\label{Transient-absorption spectra for intense pump and probe pulses}

\begin{figure*}
\includegraphics[width= \linewidth]{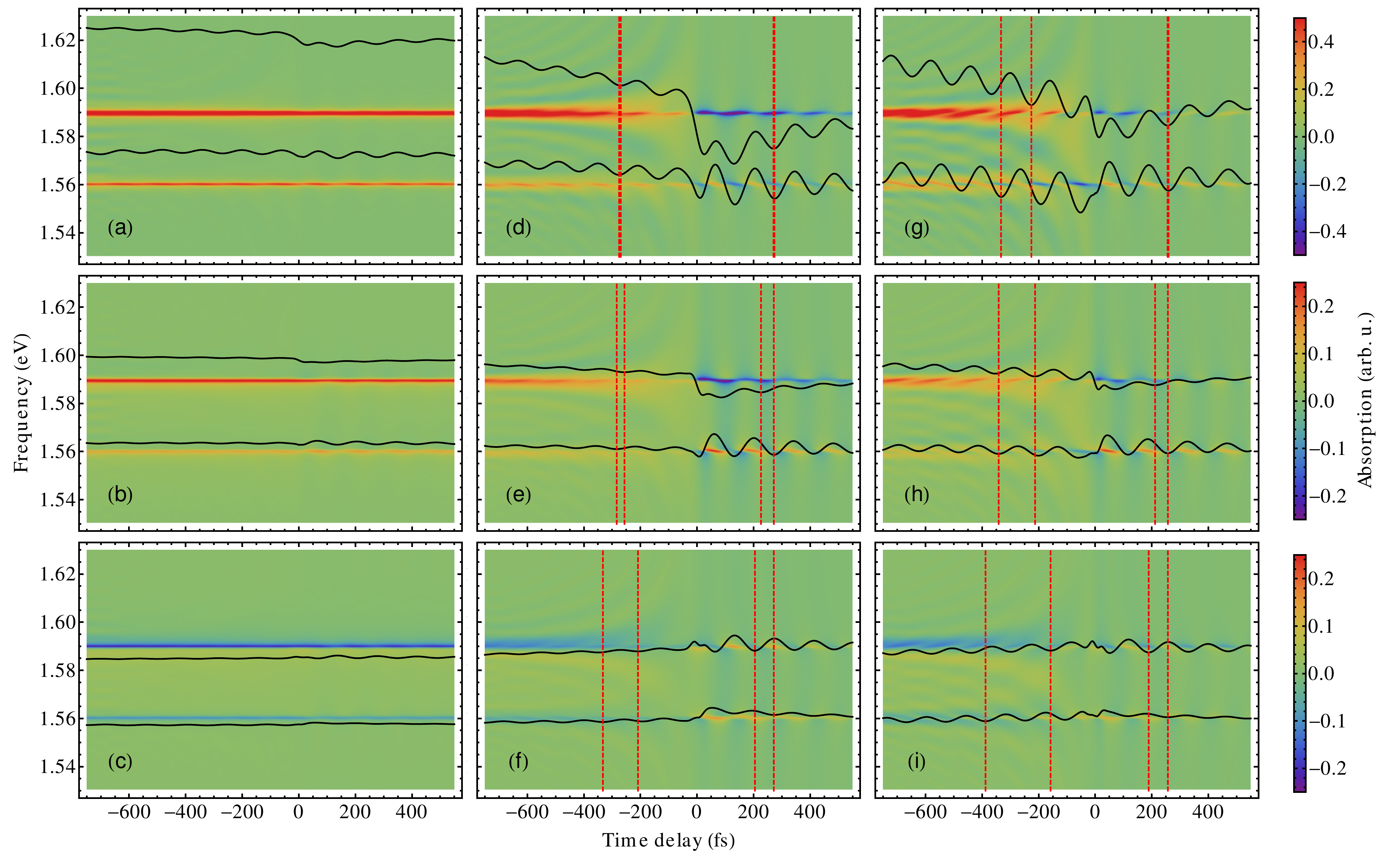}
\caption{Absorption spectra for laser frequencies of $\omega_{\mathrm{L}} = 1.59\,\mathrm{eV}$, pump intensities of [(a), (b), (c)] $I_{\mathrm{pu}}=1\times10^9\,\mathrm{W/cm^2}$, [(d), (e), (f)] $I_\mathrm{pu}=1\times10^{10}\,\mathrm{W/cm^2}$, and [(g), (h), (i)] $I_\mathrm{pu}=2.8\times10^{10}\,\mathrm{W/cm^2}$, 
and probe intensities of [(a), (d), (g)] $I_\mathrm{pr}=1\times10^9\,\mathrm{W/cm^2}$, [(b), (e), (h)] $I_\mathrm{pr}=1\times10^{10}\,\mathrm{W/cm^2}$, and [(c), (f), (i)] $I_\mathrm{pu}=2.8\times10^{10}\,\mathrm{W/cm^2}$. In each panel, the top (bottom) black lines represent the absorption spectra evaluated at the transition energy $\omega_{31}$ ($\omega_{21}$) in arbitrary units. All black lines are on the same scale, with the 0 aligned on the corresponding transition energy. The red dashed lines correspond to local minima of the spectra evaluated at $\omega=\omega_{21}$ and $\omega=\omega_{31}$.}
\label{fig:gridfreq}
\end{figure*}

Here, we apply our three-level model to study Rb atoms excited by intense femtosecond probe and pump pulses. Simulated time-delay dependent transient-absorption spectra, obtained by numerically solving Eq.~(\ref{eq:matrixC}) and then using this solution in Eqs.~(\ref{eq:spectrum1}) and (\ref{eq:spectrumaveraged}), are displayed in Fig.~\ref{fig:gridfreq} for representative values of pump- and probe-pulse intensities and for a laser frequency of $\omega_{\mathrm{L}} = 1.59\,\mathrm{eV}$. For all sets of intensities investigated, two absorption lines can be distinguished, respectively centered on the transition energies $\omega_{21} = 1.56\,\mathrm{eV}$ and $\omega_{31} = 1.59\,\mathrm{eV}$. The shape and amplitude of these lines is modulated as a function of time delay, featuring oscillations whose period of $2\pi/\omega_{32} = 140\,\mathrm{fs}$ is given by the beating frequency $\omega_{32}$. This is stressed by the black lines, showing the spectra evaluated at the two transition energies $\omega_{21}$ and $\omega_{31}$ as a function of $\tau$. 

Figures~\ref{fig:gridfreq}(a), \ref{fig:gridfreq}(b), and \ref{fig:gridfreq}(c) show transient-absorption spectra for a weak pump intensity of $I_{\mathrm{pu}} = 1\times 10^9\,\mathrm{W/cm^2}$ and three different values of probe intensity. Firstly, we notice that the amplitude of the time-delay-dependent oscillations displayed by the spectra is very small for these weak values of the pump intensity. The shape and amplitude of the absorption lines remain almost completely unchanged throughout the range of $\tau$ displayed, with no significant features distinguishing between positive and negative time delays. By modifying the probe intensity, we notice a variation in the strength of the lines, going from absorption for a weak intensity of $I_{\mathrm{pr}} = 1\times 10^9\,\mathrm{W/cm^2}$ to emission at higher values of intensity.

When higher values of pump-pulse intensity are employed, clear time-delay-dependent features can be distinguished. The amplitude and the phase of these oscillations in $\tau$ varies differently, for positive and negative time delays, as a function of pump and probe intensities. Figures~\ref{fig:gridfreq}(a), \ref{fig:gridfreq}(d), and \ref{fig:gridfreq}(g) show spectra evaluated for a weak probe intensity of $I_{\mathrm{pr}} = 1\times 10^9\,\mathrm{W/cm^2}$ and increasing values of $I_{\mathrm{pu}}$. For intermediate values of the pump-pulse intensity ($I_{\mathrm{pu}} = 1\times 10^{10}\,\mathrm{W/cm^2}$) and for both positive and negative time delays, the phase of the exhibited time-delay-dependent spectra is the same for the two transition energies, as evinced by the red dashed lines which highlight the position of the minima of $\mathcal{S}(\omega_{21},\tau)$ and $\mathcal{S}(\omega_{31},\tau)$. However, as already discussed in Ref.~\cite{PhysRevLett.115.033003}, a shift can be recognized for a higher pump intensity of $I_{\mathrm{pu}} = 2.8\times 10^{10}\,\mathrm{W/cm^2}$: while the spectra evaluated at $\omega_{21}$ and $\omega_{31}$ shift in opposite directions for $\tau<0$ as a clear and distinguishable signature of the onset of strong-field effects, a common shift in the same direction takes place at $\tau>0$ when the pump-pulse intensity is increased. 

Recognizing these strong-field-induced features and understanding them in terms of intensity-dependent atomic phases becomes more complex when a probe pulse is used which is not sufficiently weak. This appears clearly when one compares Figs.~\ref{fig:gridfreq}(d), \ref{fig:gridfreq}(e), and \ref{fig:gridfreq}(f), where results are shown for an intermediately strong pump pulse and different values of the probe intensity. Both at positive and negative time delays, absorption lines evaluated at $\omega_{21}$ and $\omega_{31}$ feature a shift in opposite directions, which becomes larger at high probe intensities. Similarly, spectra displayed in Figs.~\ref{fig:gridfreq}(g), \ref{fig:gridfreq}(h), and \ref{fig:gridfreq}(i) for a pump intensity of $I_{\mathrm{pu}} = 2.8\times 10^{10}\,\mathrm{W/cm^2}$ show that a probe-pulse-induced shift of the spectra evaluated at $\omega_{21}$ and $\omega_{31}$ arises for growing values of $I_{\mathrm{pr}}$: at negative time delays, this enlarges the already existent shift due to the strong pump pulse; for positive time delays, where the increase in $I_{\mathrm{pu}}$ causes an aligned, common shift of $\mathcal{S}(\omega_{21},\tau)$ and $\mathcal{S}(\omega_{31},\tau)$, the presence of an intense probe pulse is reflected in additional shifts, analogous to those already recognized for $I_{\mathrm{pu}} = 1\times 10^{10}\,\mathrm{W/cm^2}$. 

It should be noticed that the spectra in Figs.~\ref{fig:gridfreq}(a), \ref{fig:gridfreq}(e), and \ref{fig:gridfreq}(i) are calculated for equal pump- and probe-pulse intensities. The dynamics of the system are, therefore, perfectly symmetric with respect to $\tau$, and the system features the same time evolution when equally delayed pump and probe pulses are used, independent of their arriving order. Nevertheless, the spectra exhibited in the above listed figures are clearly not symmetric with respect to $\tau$, and different amplitudes and phases of the time-delay-dependent features of $\mathcal{S}(\omega,\tau)$ can be recognized at $\tau>0$ or $\tau<0$, in spite of identical underlying dynamics. This can be understood by noticing that the spectrum arises from the interference between the electric dipole response of the atomic system with the probe pulse: even when the quantum dynamics are identical, the spectrum still reveals how these influence the first-(second-)arriving probe pulse for $\tau<0$ ($\tau>0$). This is also evident from the definition of the absorption spectrum~(\ref{eq:spectrum1}), where the Fourier transform is always centered on the central time $\tau$ of the probe pulse, and then from the analytic models in Eqs.~(\ref{eq:Sexpprpu}) and (\ref{eq:Sexppupr}), respectively describing time-delay-averaged probe-pump and pump-probe spectra from a noncollinear geometry. Even when identical pump and probe pulses are used ($\hat{U}_{\mathrm{pr}} = \hat{U}_{\mathrm{pu}} $), the spectra evaluated at positive and negative time delays are determined by different interaction-operator matrix elements and hence differ.

\begin{figure}[t]
\includegraphics[width= \columnwidth]{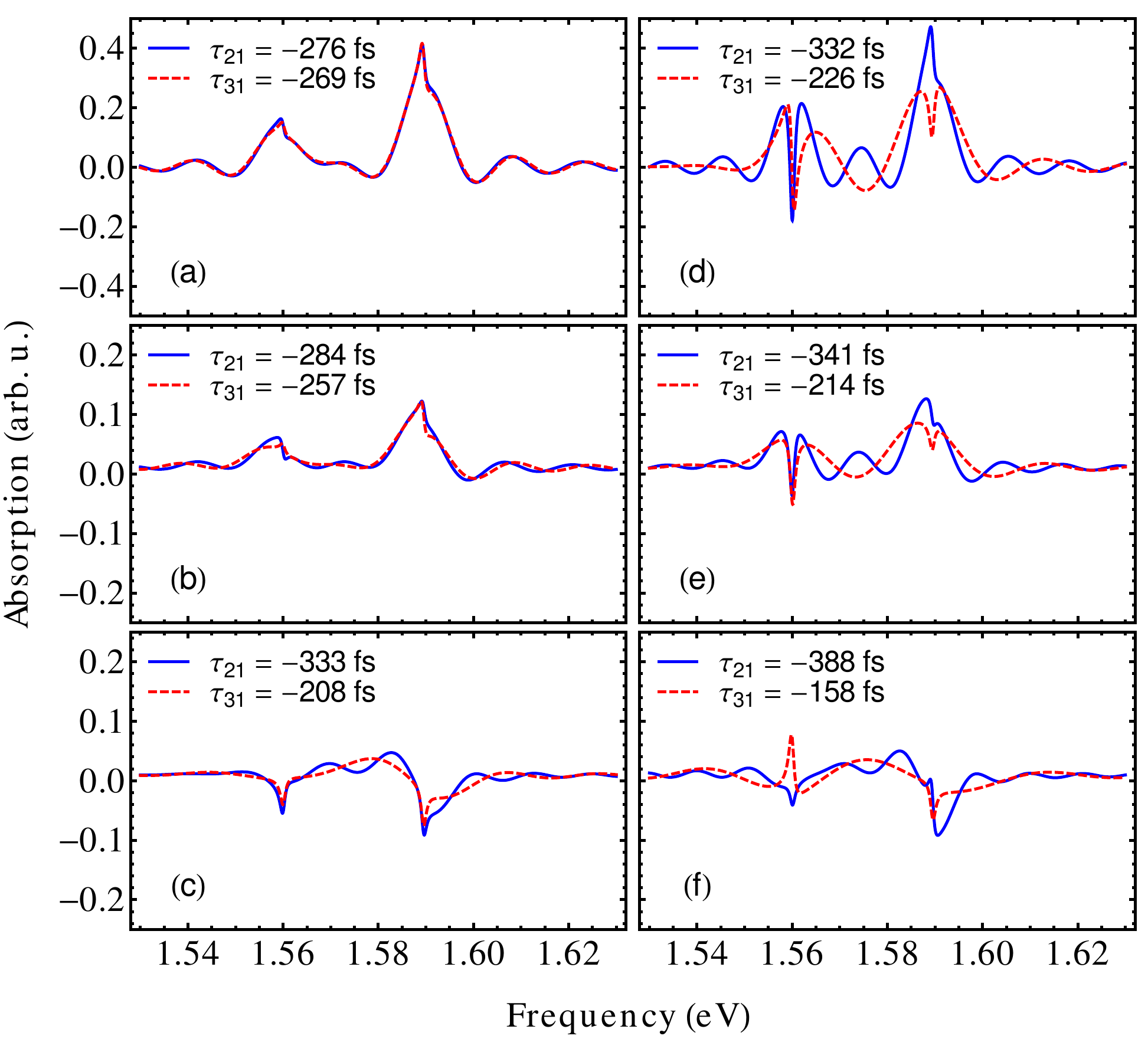}
\caption{Probe-pump transient-absorption spectra evaluated as a function of frequency at two different time delays $\tau_{21}$ (blue, continuous) and $\tau_{31}$ (red, dashed), for laser frequencies of $\omega_{\mathrm{L}} = 1.59\,\mathrm{eV}$, pump intensities of [(a), (b), (c)] $I_\mathrm{pu}=1\times10^{10}\,\mathrm{W/cm^2}$ and [(d), (e), (f)] $I_\mathrm{pu}=2.8 \times10^{10} \,\mathrm{W/cm^2}$, and probe intensities of [(a), (d)] $I_\mathrm{pr}=1\times10^9\,\mathrm{W/cm^2}$, [(b), (e)] $I_\mathrm{pr}=1\times10^{10}\,\mathrm{W/cm^2}$, and [(c), (f)] $I_\mathrm{pr}=2.8\times10^{10}\,\mathrm{W/cm^2}$. For each panel, the time delay $\tau_{21}$ ($\tau_{31}$) at which the spectrum is evaluated is associated with the local-minimum point of $\mathcal{S}(\omega_{21},\tau)$ [$\mathcal{S}(\omega_{31},\tau)$] highlighted in Fig.~\ref{fig:gridfreq} by a red, dashed line for $\tau<0$.}
\label{fig:tauminPRPU}
\end{figure}

\begin{figure}[t]
\includegraphics[width= \columnwidth]{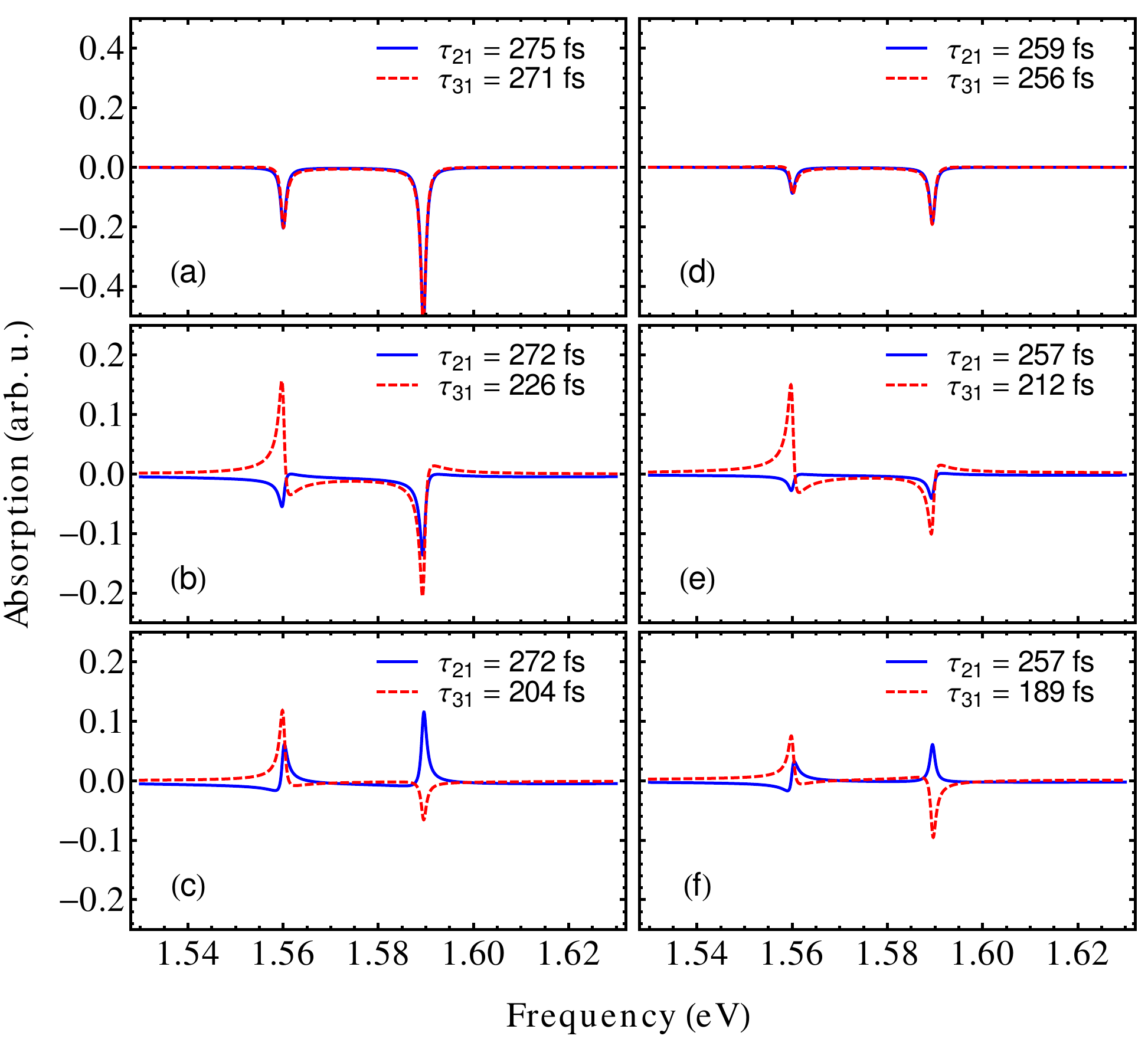}
\caption{Pump-probe transient-absorption spectra evaluated as a function of frequency at two different time delays $\tau_{21}$ (blue, continuous) and $\tau_{31}$ (red, dashed), for the same parameters used in Fig.~\ref{fig:tauminPRPU}. For each panel, the time delay $\tau_{21}$ ($\tau_{31}$) at which the spectrum is evaluated is associated with the local-minimum point of $\mathcal{S}(\omega_{21},\tau)$ [$\mathcal{S}(\omega_{31},\tau)$] highlighted in Fig.~\ref{fig:gridfreq} by a red, dashed line for $\tau>0$.}
\label{fig:tauminPUPR}
\end{figure}

In the previous discussion we have focused on the time-delay-dependent properties of the spectra $\mathcal{S}(\omega_{k1}, \tau)$, evaluated at the transition energies $\omega_{k1}$. However, the identification of $\omega_{21}$ and $\omega_{31}$ may not be straightforward experimentally, affecting the properties of the observed time-delay-dependent features and the quantification of the associated phases. In order to better discuss this point and describe the line-shape changes ensuing from the presence of intense pump and probe pulses, in Figs.~\ref{fig:tauminPRPU} and \ref{fig:tauminPUPR}, for a probe-pump and pump-probe setup, respectively, we present transient-absorption spectra $\mathcal{S}(\omega, \tau_{k1})$, $k \in\{2,\,3\}$, evaluated as a function of frequency for fixed values of the time delay, $\tau_{21}$ and $\tau_{31}$. Here, the time delay $\tau_{21}$ ($\tau_{31}$) is the one for which $\mathcal{S}(\omega_{21},\tau)$ [$\mathcal{S}(\omega_{31},\tau)$] has a local minimum, as identified in Fig.~\ref{fig:gridfreq} by the red, dashed lines. The pictures show that the identified local-minimum points are not necessarily associated with emission peaks pointing downwards. Furthermore, for negative time delays, where additional frequency modulations appear as shown in Figs.~\ref{fig:gridfreq} and \ref{fig:tauminPRPU}, one has to disentangle the behavior of the peaks centered on $\omega_{k1}$ from the remaining modulations appearing as a function of frequency. Nevertheless, all panels confirm that it is possible to isolate the time-delay-dependent behavior of this central peak and, thereby, identify the particular time delay at which this is minimal. 

\begin{figure*}[t]
\includegraphics[width= \linewidth]{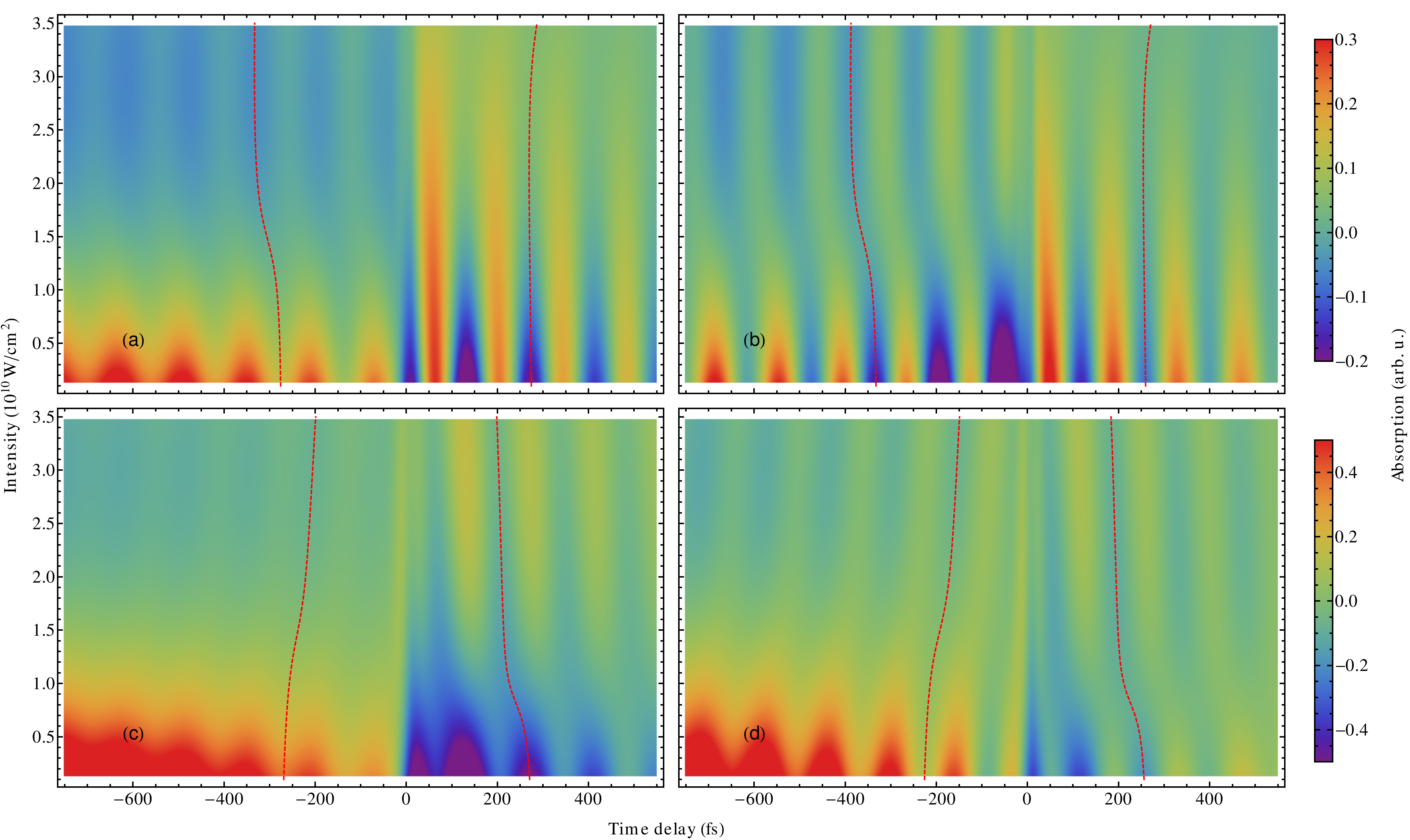}
\caption{Absorption spectra evaluated at [(a), (b)] $\omega_{21}$ and [(c), (d)] $\omega_{31}$ as a function
of probe intensity and time delay, for a laser frequency of $\omega_{\mathrm{L}} = 1.59\,\mathrm{eV}$ and at fixed pump intensities of [(a), (c)] $I_\mathrm{pu}=1\times10^{10}\,\mathrm{W/cm^2}$ and [(b), (d)] $I_\mathrm{pu}=2.8\times10^{10}\,\mathrm{W/cm^2}$. The red dashed lines correspond to the local-minimum points (as a function of probe-pulse intensity) of $\mathcal{S}(\omega_{21},\tau)$ or $\mathcal{S}(\omega_{31},\tau)$.
}
\label{fig:gridInt}
\end{figure*}

Encouraged by the results displayed in Figs.~\ref{fig:tauminPRPU} and \ref{fig:tauminPUPR}, in the following we focus on $\mathcal{S}(\omega_{k1}, \tau)$ and the corresponding time-delay-dependent oscillations in order to draw conclusions about strong-field-induced atomic phases. Figure~\ref{fig:gridInt} shows the amplitude of the numerically calculated spectra $\mathcal{S}(\omega_{21},\tau)$ and $\mathcal{S}(\omega_{31},\tau)$ as a function of probe-pulse intensity for two different values of $I_{\mathrm{pu}}$. The shifts in the phase of the time-delay-dependent spectra is here clearly apparent. For $\tau>0$ or $\tau<0$, the effect of the intense pump and probe pulses appears in the spectrum as independent pump- and probe-induced phase shifts. In the following, in order to investigate this point further and identify how atomic phase changes are encoded in transient-absorption spectra, we interpret our results in terms of the interaction operators introduced in Subsec.~\ref{Analytical model in terms of interaction operators}.

\subsection{Interpretation of pump- and probe-pulse-induced phases in terms of interaction-operator matrix elements}

\label{Interpretation of pump- and probe-pulse-induced phases in terms of interaction-operator matrix elements}

Here, we use Eqs.~(\ref{eq:Sexpprpu}) and (\ref{eq:Sexppupr}) in order to interpret the numerically calculated transient-absorption spectra presented in Subsec.~\ref{Transient-absorption spectra for intense pump and probe pulses} in terms of interaction-operator matrix elements. In particular, we focus on the phase of the time-delay-dependent oscillations exhibited by $\mathcal{S}(\omega_{21},\tau)$ and $\mathcal{S}(\omega_{31},\tau)$ [Fig.~\ref{fig:gridInt}], and show how these can be understood via the strong-field-induced atomic phases quantified in $\hat{U}_{\mathrm{pu}}$ and $\hat{U}_{\mathrm{pr}}$. For both a probe-pump and a pump-probe setup, we develop analytical interpretation models, calculate $\hat{U}_{\mathrm{pu}}$ and $\hat{U}_{\mathrm{pr}}$ with Eqs.~(\ref{eq:matrixU0}) and (\ref{eq:definitionU}), and then use these interpretation models to understand the phase features displayed by the transient-absorption spectra in Figs.~\ref{fig:gridfreq} and \ref{fig:gridInt}. Finally, we further investigate the dependence of the phases extractable from transient-absorption spectra upon the laser frequency of the pump and probe pulses.

\subsubsection{Probe-pump setup}

Firstly, we focus on the probe-pump interpretation model given by Eq.~(\ref{eq:Sexpprpu}), aiming at better understanding the properties of the spectrum evaluated at $\omega = \omega_{k1}$. For interpretation purposes, since $\omega_{32} \gg \gamma_k$, we are allowed to neglect in first approximation the term proportional to
$D_{1k'}^*/( \uimm\,\omega_{kk'} +\gamma_k/2)$, with $k'\in\{2,\,3\}$, $k'\neq k$, $\omega_{kk'} = \pm \omega_{32}$, thus obtaining
\begin{equation}
\begin{split}
\mathcal{S}_{\mathrm{prpu}}(\omega_{k1},\tau) 
& \propto-\frac{\omega}{K_{\mathrm{pr}}}\Imm\biggl\{2\frac{D_{1k}^*}{ \gamma_k} \\
&\ \ \times \bigl[U_{\mathrm{pr},11}U_{\mathrm{pr},k1}^*(1-\eu^{\frac{\gamma_k}{2}\tau})\\
&\ \ \ + U_{\mathrm{pu},11}U_{\mathrm{pu},kk}^* U_{\mathrm{pr},11} U_{\mathrm{pr},k1}^*\eu^{\frac{\gamma_k}{2} \tau}\\
& \ \ \ + U_{\mathrm{pu},11}U_{\mathrm{pu},kk'}^* U_{\mathrm{pr},11} U_{\mathrm{pr},k'1}^* \eu^{\uimm \omega_{kk'} \tau} \eu^{\frac{\gamma_{k'}}{2} \tau}\bigr]\biggr\}.
\end{split}
\end{equation}
The only term which displays oscillations as a function of $\tau$ is given by
\begin{equation}
\begin{split}
\tilde{\mathcal{S}}_{\mathrm{prpu}}(\omega_{k1},\tau) 
& \propto-2\frac{\omega}{K_{\mathrm{pr}}}\frac{D_{1k}}{ \gamma_k}\eu^{\frac{\gamma_{k'}}{2} \tau} \Imm\bigl(Y_{\mathrm{pu},k}\,Y_{\mathrm{pr},k}\,\eu^{\uimm \omega_{kk'} \tau } \bigr),
\end{split}
\end{equation}
with 
\begin{equation}
\begin{aligned}
Y_{\mathrm{pr},k} &= U_{\mathrm{pr},11} U_{\mathrm{pr},k'1}^*,\\
Y_{\mathrm{pu},k} &= U_{\mathrm{pu},11} U_{\mathrm{pu},kk'}^*,
\end{aligned}
\end{equation}
and where we have used explicitly the fact that, for our atomic implementation with Rb atoms, the projections $D_{1k}$ of the dipole-moment matrix elements $\boldsymbol{D}_{1k}$ along the pulse polarization axis $\hat{\boldsymbol{e}}_z$ are real. We can more explicitly write
\begin{equation}
\begin{split}
\tilde{\mathcal{S}}_{\mathrm{prpu}}(\omega_{21},\tau) 
& \propto-2\frac{\omega}{K_{\mathrm{pr}}}\frac{D_{12}}{ \gamma_2}\eu^{\frac{\gamma_3}{2} \tau}\,|Y_{\mathrm{pu},2}|\,|Y_{\mathrm{pr},2}| \\
&\ \times \Imm\bigl\{\eu^{-\uimm [\omega_{32} \tau - \arg{(Y_{\mathrm{pu},2})} - \arg{(Y_{\mathrm{pr},2})}] } \bigr\}\\
& =-2\frac{\omega}{K_{\mathrm{pr}}}\frac{D_{12}}{ \gamma_2}\eu^{\frac{\gamma_3}{2} \tau}\,|Y_{\mathrm{pu},2}|\,|Y_{\mathrm{pr},2}|  \\
&\ \times \sin{ [\omega_{32} \tau - \arg{(Y_{\mathrm{pu},2})} - \pi - \arg{(Y_{\mathrm{pr},2})} ] }
\end{split}
\end{equation}
and
\begin{equation}
\begin{split}
\tilde{\mathcal{S}}_{\mathrm{prpu}}(\omega_{31},\tau) 
& \propto-2\frac{\omega}{K_{\mathrm{pr}}}\frac{D_{13}}{ \gamma_3}\eu^{\frac{\gamma_2}{2} \tau}\,|Y_{\mathrm{pu},3}|\,|Y_{\mathrm{pr},3}|\\
&\ \times \Imm\bigl\{\eu^{\uimm [\omega_{32} \tau + \arg{(Y_{\mathrm{pu},3})} + \arg{(Y_{\mathrm{pr},3})}] } \bigr\}\\
& =-2\frac{\omega}{K_{\mathrm{pr}}}\frac{D_{13}}{ \gamma_3}\eu^{\frac{\gamma_2}{2} \tau}\,|Y_{\mathrm{pu},3}|\,|Y_{\mathrm{pr},3}| \\
&\ \times  \sin{ [\omega_{32} \tau + \arg{(Y_{\mathrm{pu},3})} + \arg{(Y_{\mathrm{pr},3})}] }.
\end{split}
\end{equation}
With
\begin{equation}
\begin{aligned}
Y_{\mathrm{pr},2} &= U_{\mathrm{pr},11} U_{\mathrm{pr},31}^*,\\
Y_{\mathrm{pr},3} &= U_{\mathrm{pr},11} U_{\mathrm{pr},21}^*,\\
Y_{\mathrm{pu},2} &= U_{\mathrm{pu},11} U_{\mathrm{pu},23}^*,\\
Y_{\mathrm{pu},3} &= U_{\mathrm{pu},11} U_{\mathrm{pu},32}^*,\\
\end{aligned}
\label{eq:Y}
\end{equation}
and the phases
\begin{equation}
\begin{aligned}
\varphi_{\mathrm{pr},2} &= -\pi -\arg{(U_{\mathrm{pr},11} U_{\mathrm{pr},31}^*)},\\
\varphi_{\mathrm{pr},3} &= \arg{(U_{\mathrm{pr},11} U_{\mathrm{pr},21}^*)},\\
\varphi_{\mathrm{pu},2} &= -\arg{(U_{\mathrm{pu},11} U_{\mathrm{pu},23}^*)},\\
\varphi_{\mathrm{pu},3} &= \arg{(U_{\mathrm{pu},11} U_{\mathrm{pu},32}^*)},\\
\end{aligned}
\label{eq:phi}
\end{equation}
this reduces to
\begin{equation}
\begin{split}
\tilde{\mathcal{S}}_{\mathrm{prpu}}(\omega_{21},\tau) 
& =-2\frac{\omega}{K_{\mathrm{pr}}}\frac{D_{12}}{ \gamma_2}\eu^{\frac{\gamma_3}{2} \tau}\,|Y_{\mathrm{pu},2}|\,|Y_{\mathrm{pr},2}|  \\
&\ \times \sin{ [\omega_{32} \tau +\varphi_{\mathrm{pr},2} + \varphi_{\mathrm{pu},2} ] }
\end{split}
\label{eq:Sprpu21}
\end{equation}
and
\begin{equation}
\begin{split}
\tilde{\mathcal{S}}_{\mathrm{prpu}}(\omega_{31},\tau) 
& =-2\frac{\omega}{K_{\mathrm{pr}}}\frac{D_{13}}{ \gamma_3}\eu^{\frac{\gamma_2}{2} \tau}\,|Y_{\mathrm{pu},3}|\,|Y_{\mathrm{pr},3}| \\
&\ \times  \sin{ [\omega_{32} \tau +\varphi_{\mathrm{pr},3} + \varphi_{\mathrm{pu},3} ]}.
\end{split}
\label{eq:Sprpu31}
\end{equation}
The intensity-dependent position of the minima of $\mathcal{S}(\omega_{k1},\tau)$ for $\tau<0$, shown in Fig.~\ref{fig:gridInt} by the red dashed lines at negative time delays, can hence be quantified via Eqs.~(\ref{eq:Sprpu21}) and (\ref{eq:Sprpu31}) in terms of $\varphi_{\mathrm{pr},k}$ and $\varphi_{\mathrm{pu},k}$. The sine functions appearing therein have local minima, respectively centered around
\begin{equation}
\begin{aligned} 
&\tau_{21}=\tau_0 - \frac{(\varphi_{\mathrm{pr,2}}+\varphi_{\mathrm{pu,2}})}{\omega_{32}},   & \text{for $\omega = \omega_{21}$, $\tau<0$}, \\
&\tau_{31}=\tau_0 - \frac{(\varphi_{\mathrm{pr,3}}+\varphi_{\mathrm{pu,3}})}{\omega_{32}},   & \text{for $\omega = \omega_{31}$, $\tau<0$}, \\
\end{aligned}
\label{eq:tau-prpu}
\end{equation}
with the additive offset $\tau_0 = -9\pi/(2\omega_{32})$. For real, positive dipole-moment matrix elements $D_{1k}$, and hence real positive pulse areas $\vartheta_k$, the intensity-dependent variables $Y_{\mathrm{pr},k}$ and $Y_{\mathrm{pu},k}$ can be explicitly written in the case of weak pulses via Eq.~(\ref{eq:weak}) as
\begin{equation}
\begin{aligned}
Y_{\mathrm{pr},2}^{\mathrm{weak}} &= -\uimm\frac{\vartheta_3}{2},\\
Y_{\mathrm{pr},3}^{\mathrm{weak}} &= -\uimm\frac{\vartheta_2}{2},\\
Y_{\mathrm{pu},2}^{\mathrm{weak}} &= -\vartheta_2\vartheta_3,\\
Y_{\mathrm{pu},3}^{\mathrm{weak}} &= -\vartheta_2\vartheta_3,\\
\end{aligned}
\label{eq:Yweak}
\end{equation}
along with the associated phases
\begin{equation}
\begin{aligned}
\varphi_{\mathrm{pr},2}^{\mathrm{weak}} &= - \pi/2,\\
\varphi_{\mathrm{pr},3}^{\mathrm{weak}} &= - \pi/2,\\
\varphi_{\mathrm{pu},2}^{\mathrm{weak}} &= \pm \pi,\\
\varphi_{\mathrm{pu},3}^{\mathrm{weak}} &= \mp \pi.\\
\end{aligned}
\label{eq:phiweak}
\end{equation}
For low intensities, the effect of the probe pulse is linearly proportional to the pulse areas $\vartheta_k$ and, therefore, of first order in the amplitude of the electric field, whereas the action of the pump pulse depends on the product of $\vartheta_2\vartheta_3$ and is hence of second order. This explains the small, almost vanishing amplitude of the time-delay-dependent oscillations displayed for $\tau<0$ by the transient-absorption spectra in Figs.~\ref{fig:gridfreq}(a), \ref{fig:gridfreq}(b), and \ref{fig:gridfreq}(c), for a small pump-pulse intensity of $I_{\mathrm{pu}} = 1\times 10^9\,\mathrm{W/cm^2}$.

\begin{figure*}[t]
\includegraphics[width= \linewidth]{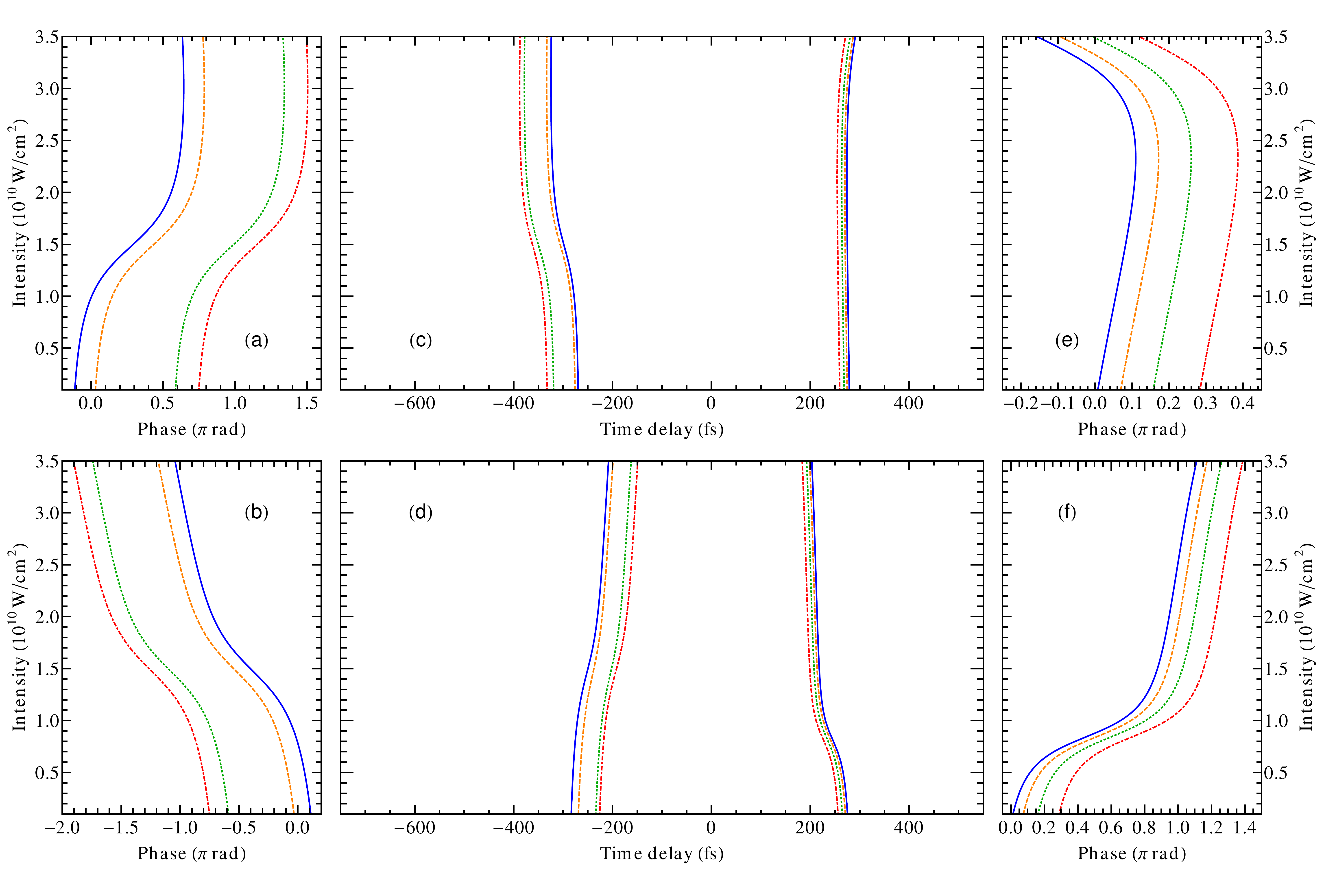}
\caption{Correspondence between strong-field-induced atomic phases and time-delay-dependent oscillations of the transient-absorption spectra for a laser frequency of $\omega_{\mathrm{L}} = 1.59\,\mathrm{eV}$. [(c), (d)] Time delays (as a function of probe pulse intensity) associated with minima in the absorption spectra (c) $\mathcal{S}(\omega_{21},\tau)$ and (d) $\mathcal{S}(\omega_{31},\tau)$ for both positive and negative time delays; [(a), (b)] associated total phases for a probe-pump setup (a) $[\varphi_{\mathrm{pr,2}}+\varphi_{\mathrm{pu,2}}-(\varphi_{\mathrm{pr,2}}^{\mathrm{weak}}+\varphi_{\mathrm{pu,2}}^{\mathrm{weak}})]$ and (b) $[\varphi_{\mathrm{pr,3}}+\varphi_{\mathrm{pu,3}}-(\varphi_{\mathrm{pr,3}}^{\mathrm{weak}}+\varphi_{\mathrm{pu,3}}^{\mathrm{weak}})]$; and [(e), (f)] associated total phases for a pump-probe setup (e)
$[\psi_{\mathrm{pr,2}}+\psi_{\mathrm{pu}}-(\psi_{\mathrm{pr,2}}^{\mathrm{weak}}+\psi_{\mathrm{pu}}^{\mathrm{weak}})]$ and (f) $[\psi_{\mathrm{pr,3}}+\psi_{\mathrm{pu}}-(\psi_{\mathrm{pr,3}}^{\mathrm{weak}}+\psi_{\mathrm{pu}}^{\mathrm{weak}})]$. In all panels, curves are displayed for pump intensities of
$I_{\mathrm{pu}}=1\times10^9\,\mathrm{W/cm^2}$ (blue continuous line), 
$I_{\mathrm{pu}}=1\times10^{10}\,\mathrm{W/cm^2}$ (orange dashed line), 
$I_{\mathrm{pu}}=1.9\times10^{10}\,\mathrm{W/cm^2}$ (green dotted line), and
$I_{\mathrm{pu}}=2.8\times10^{10}\,\mathrm{W/cm^2}$ (red dashed-dotted line).
}
\label{fig:minimaphases}
\end{figure*}

In Figs.~\ref{fig:minimaphases}(a) and \ref{fig:minimaphases}(b), the total phases $[\varphi_{\mathrm{pr},2} + \varphi_{\mathrm{pu},2} - (\varphi_{\mathrm{pr},2}^{\mathrm{weak}} + \varphi_{\mathrm{pu},2}^{\mathrm{weak}})]$ and $[\varphi_{\mathrm{pr},3} + \varphi_{\mathrm{pu},3} - (\varphi_{\mathrm{pr},3}^{\mathrm{weak}} + \varphi_{\mathrm{pu},3}^{\mathrm{weak}})]$ [Eqs.~(\ref{eq:phi}) and (\ref{eq:phiweak}) after numerical calculation of $\hat{U}_{\mathrm{pr}}$ and $\hat{U}_{\mathrm{pu}}$ via Eqs.~(\ref{eq:matrixU0}) and (\ref{eq:definitionU})] are exhibited, as a function of $I_{\mathrm{pr}}$ and for a discrete set of values of $I_{\mathrm{pu}}$. The very good agreement between the intensity dependence of these phases and the shift displayed by the time-delay-dependent features of $\mathcal{S}(\omega_{21}, \tau)$ and $\mathcal{S}(\omega_{31}, \tau)$ [Fig.~\ref{fig:gridInt} and Figs.~\ref{fig:minimaphases}(c) and \ref{fig:minimaphases}(d) at negative time delays] confirms the validity of our analytical interpretation model and in particular of Eq.~(\ref{eq:tau-prpu}). The shift in the phases [Figs.~\ref{fig:minimaphases}(a) and \ref{fig:minimaphases}(b)] is reflected by an oppositely directed shift in the local-minimum points [Figs.~\ref{fig:minimaphases}(c) and \ref{fig:minimaphases}(d)] as a function of $I_{\mathrm{pr}}$ and $I_{\mathrm{pu}}$, as expected from the minus sign in Eq.~(\ref{eq:tau-prpu}).

In order to understand the physics underlying the phase shifts $\varphi_{\mathrm{pr},k}$ appearing in the spectrum, we can use the schematic illustration of $\hat{U}(I)$ in Fig.~\ref{fig:schematic-explanation} to clarify the meaning of the terms appearing in Eqs.~(\ref{eq:Y}) and (\ref{eq:phi}). The associated terms $Y_{\mathrm{pr},k} =U_{\mathrm{pr},11} U_{\mathrm{pr},k'1}^* $, $k'\neq k$, are the coherences (in amplitude and phase) generated by the first-arriving probe pulse acting on the ground state. The shift displayed by $\mathcal{S}(\omega_{k1},\tau)$ is therefore related to the phase of these strong-field-induced coherences. The different sign appearing in the definition of $\varphi_{\mathrm{pr},2}$ and $\varphi_{\mathrm{pr},3}$ also explains why the time-delay-dependent oscillations of $\mathcal{S}(\omega_{21},\tau)$ and $\mathcal{S}(\omega_{31},\tau)$ shift in opposite directions for increasing probe-pulse intensities [Fig.~\ref{fig:minimaphases}(c) and \ref{fig:minimaphases}(d)].

The second-arriving intense pump pulse nonlinearly modifies an already existent superposition of excited states. The shifts $\varphi_{\mathrm{pu},2}$ and $\varphi_{\mathrm{pu},3}$ in the oscillating features of $\mathcal{S}(\omega_{21},\tau)$ and $\mathcal{S}(\omega_{31},\tau)$, respectively, quantify the changes in the atomic phases induced by the pump pulse. This can be recognized via inspection of the associated interaction-operator matrix elements, $Y_{\mathrm{pu},k} = U_{\mathrm{pu},11} U_{\mathrm{pu},kk'}^*$, $k'\neq k$ [Eqs.~(\ref{eq:Y}) and (\ref{eq:phi})], which describe how the pump pulse transforms an initial coherence between ground state and excited state $|k'\rangle$ into a final coherence between ground state and excited state $|k\rangle$ (see also the schematic illustration in Fig.~\ref{fig:schematic-explanation}). The ensuing phase change determines the shift appearing in the oscillating features of the transient-absorption spectrum. Also in this case, the shift in opposite directions displayed by $\mathcal{S}(\omega_{21},\tau)$ and $\mathcal{S}(\omega_{31},\tau)$ for rising values of $I_{\mathrm{pu}}$ [Fig.~\ref{fig:minimaphases}(c) and \ref{fig:minimaphases}(d)] is a consequence of the opposite sign with which $\varphi_{\mathrm{pu},2}$ and $\varphi_{\mathrm{pu},3}$ are related to the interaction-operator matrix elements [Eq.~(\ref{eq:phi})].

\subsubsection{Pump-probe setup}

Here, we focus on the positive-time-delay part of the spectrum, and use the associated interpretation model given by Eq.~(\ref{eq:Sexppupr}) in order to better understand the properties of the spectra evaluated at $\omega = \omega_{k1}$. For this purpose, as already performed in the previous part, we can neglect terms given by
$D_{1k'}^*/( \uimm \omega_{kk'}+ \gamma_{k'}/2)$ in Eq.~(\ref{eq:Sexppupr}), and thus identify those contributions which are responsible for the oscillations exhibited by the spectrum as a function of $\tau$:
\begin{equation}
\begin{split}
&\tilde{\mathcal{S}}_{\mathrm{pupr}}(\omega_{k1},\tau) \propto -2\frac{\omega}{K_{\mathrm{pr}}}\frac{D_{1k}}{ \gamma_k}\, \eu^{-\frac{\gamma_2+\gamma_3}{2}\tau} \\
&\ \ \ \ \ \ \times \Imm\bigl(U_{\mathrm{pr},12} U_{\mathrm{pr},k3}^* U_{\mathrm{pu,}21} U^*_{\mathrm{pu,}31} \,\eu^{\uimm\omega_{32}\tau}\\
&\ \ \ \ \ \ \ \ \  +U_{\mathrm{pr},13} U_{\mathrm{pr},k2}^* U_{\mathrm{pu,}31} U^*_{\mathrm{pu,}21} \,\eu^{-\uimm\omega_{32}\tau} \bigr).
\end{split}
\label{eq:Sexppupr-justonepart}
\end{equation}
Also in this case, we have used explicitly the fact that the dipole-moment matrix elements $D_{1k}$ are real. By introducing the intensity-dependent pump and probe variables
\begin{equation}
\begin{aligned}
Z_{\mathrm{pu}} &= U_{\mathrm{pu,}21} U^*_{\mathrm{pu,}31},\\
A_{\mathrm{pr},k} &= U_{\mathrm{pr},12} U_{\mathrm{pr},k3}^*,\\
B_{\mathrm{pr},k} &= U_{\mathrm{pr},13} U_{\mathrm{pr},k2}^*,\\
\end{aligned}
\label{eq:ZAB}
\end{equation}
we can write Eq.~(\ref{eq:Sexppupr-justonepart}) as
\begin{equation}
\begin{split}
&\tilde{\mathcal{S}}_{\mathrm{pupr}}(\omega_{k1},\tau) \propto -2\frac{\omega}{K_{\mathrm{pr}}}\frac{D_{1k}}{ \gamma_k}\, \eu^{-\frac{\gamma_2+\gamma_3}{2}\tau} \\
&\ \ \ \ \ \ \times \Imm\bigl(A_{\mathrm{pr},k} \,Z_{\mathrm{pu}} \,\eu^{\uimm\omega_{32}\tau} + B_{\mathrm{pr},k}\, Z_{\mathrm{pu}}^*  \,\eu^{-\uimm\omega_{32}\tau} \bigr),
\end{split}
\end{equation}
and observe that the pump pulse equally acts on both terms of the above sums, resulting in a phase shift 
\begin{equation}
\psi_{\mathrm{pu}} = \arg{(Z_{\mathrm{pu}})} = \arg{(U_{\mathrm{pu,}21} U^*_{\mathrm{pu,}31})}.
\label{eq:psipu}
\end{equation}
Furthermore, since $\Imm(z) = - \Imm(z^*)$, we have that $\Imm \{B_{\mathrm{pr},k}\, 
Z_{\mathrm{pu}}^*  \,\eu^{\uimm\omega_{32}\tau} \} = - \Imm \{B_{\mathrm{pr},k}^*\, 
Z_{\mathrm{pu}} \,\eu^{\uimm\omega_{32}\tau} \}$, and hence
\begin{equation}
\begin{split}
&\tilde{\mathcal{S}}_{\mathrm{pupr}}(\omega_{k1},\tau) \propto -2\frac{\omega}{K_{\mathrm{pr}}}\frac{D_{1k}}{ \gamma_k}\, \eu^{-\frac{\gamma_2+\gamma_3}{2}\tau} \,|Z_{\mathrm{pu}}|\\
&\ \ \ \ \ \ \times \Imm\bigl[(A_{\mathrm{pr},k} - B_{\mathrm{pr},k}^*) \,\eu^{\uimm(\omega_{32}\tau + \psi_{\mathrm{pu}} )} \bigr].
\end{split}
\end{equation}
By further introducing the phases
\begin{equation}
\begin{split}
\psi_{\mathrm{pr},2} &= \arg{(A_{\mathrm{pr},2} - B_{\mathrm{pr},2}^*)} \\
&= \arg{(U_{\mathrm{pr},12} U_{\mathrm{pr},23}^*- U_{\mathrm{pr},22} U_{\mathrm{pr},13}^*)},\\
\psi_{\mathrm{pr},3} &= \arg{(A_{\mathrm{pr},3} - B_{\mathrm{pr},3}^*)} \\
&= \arg{(U_{\mathrm{pr},12} U_{\mathrm{pr},33}^*- U_{\mathrm{pr},32} U_{\mathrm{pr},13}^*)},
\end{split}
\label{eq:psipr}
\end{equation}
the spectrum can be written as
\begin{equation}
\begin{split}
&\tilde{\mathcal{S}}_{\mathrm{pupr}}(\omega_{k1},\tau) \propto -2\frac{\omega}{K_{\mathrm{pr}}}\frac{D_{1k}}{ \gamma_k}\, \eu^{-\frac{\gamma_2+\gamma_3}{2}\tau} \,|Z_{\mathrm{pu}}|\, |A_{\mathrm{pr},k} - B_{\mathrm{pr},k}^*|\\
&\ \ \ \ \ \ \times \sin{(\omega_{32}\tau + \psi_{\mathrm{pu}} + \psi_{\mathrm{pr},k} ) }.
\end{split}
\label{eq:Spupr}
\end{equation}
This implies that the intensity-dependent positions of the minima of $\mathcal{S}(\omega_{k1},\tau)$, shown in Fig.~\ref{fig:gridInt} by the red dashed lines at positive time delays, can be quantified via Eq.~(\ref{eq:Spupr}) in terms of $\psi_{\mathrm{pu}}$ and $\psi_{\mathrm{pr},k}$. The sine functions appearing therein have local minima respectively centered around
\begin{equation}
\begin{aligned} 
&\tau_{21} = \tau_0 - \frac{(\psi_{\mathrm{pu}}+\psi_{\mathrm{pr,2}})}{\omega_{32}},   & \text{for $\omega = \omega_{21}$, $\tau>0$,} \\
&\tau_{31} = \tau_0 - \frac{(\psi_{\mathrm{pu}}+\psi_{\mathrm{pr,3}})}{\omega_{32}},   & \text{for $\omega = \omega_{31}$, $\tau>0$,} \\
\end{aligned}
\label{eq:tau-pupr}
\end{equation}
with the additive offset $\tau_0 = 7\pi/(2\omega_{32})$. For real, positive dipole-moment matrix elements $D_{1k}$, and hence real positive pulse areas $\vartheta_k$, the intensity-dependent variables $Z_{\mathrm{pu}}$ and $|A_{\mathrm{pr},k} - B_{\mathrm{pr},k}^*|$ can be explicitly written in the case of weak pulses via Eq.~(\ref{eq:weak}) as
\begin{equation}
\begin{aligned}
Z_{\mathrm{pu}}^{\mathrm{weak}} &= \frac{\vartheta_2\vartheta_3}{4},\\
A_{\mathrm{pr},2}^{\mathrm{weak}} - (B_{\mathrm{pr},2}^{\mathrm{weak}})^*  &= \uimm\frac{\vartheta_3}{2},\\
A_{\mathrm{pr},3}^{\mathrm{weak}} - (B_{\mathrm{pr},3}^{\mathrm{weak}})^*  &= \uimm\frac{\vartheta_2}{2},
\end{aligned}
\label{eq:ZABweak}
\end{equation}
along with the associated phases
\begin{equation}
\begin{aligned}
\psi_{\mathrm{pu}}^{\mathrm{weak}} &= 0,\\
\psi_{\mathrm{pr},2}^{\mathrm{weak}} &= \pi/2,\\
\psi_{\mathrm{pr},3}^{\mathrm{weak}} &= \pi/2.\\
\end{aligned}
\label{eq:psiweak}
\end{equation}
Also in a pump-probe setup, the effect of a weak probe pulse is linearly proportional to the pulse areas $\vartheta_k$ and, therefore, of first order in the amplitude of the electric field. The action of a weak pump pulse depends on the product of $\vartheta_2\vartheta_3$ and is hence of second order. Also in this case, this explains the small, almost vanishing amplitude of the time-delay-dependent oscillations displayed for $\tau>0$ by the transient-absorption spectra in Figs.~\ref{fig:gridfreq}(a), \ref{fig:gridfreq}(b), and \ref{fig:gridfreq}(c), for a small pump-pulse intensity of $I_{\mathrm{pu}} = 1\times 10^9\,\mathrm{W/cm^2}$.

Figures~\ref{fig:minimaphases}(e) and \ref{fig:minimaphases}(f) display the total phases $[\psi_{\mathrm{pu}} + \psi_{\mathrm{pr},2} - (\psi_{\mathrm{pu}}^{\mathrm{weak}} + \psi_{\mathrm{pr},2}^{\mathrm{weak}})]$ and $[\psi_{\mathrm{pu}} + \psi_{\mathrm{pr},3} - (\psi_{\mathrm{pu}}^{\mathrm{weak}} + \psi_{\mathrm{pr},3}^{\mathrm{weak}})]$ [Eqs.~(\ref{eq:psipu}), (\ref{eq:psipr}), and (\ref{eq:psiweak}), after numerical calculation of $\hat{U}_{\mathrm{pr}}$ and $\hat{U}_{\mathrm{pu}}$ via Eqs.~(\ref{eq:matrixU0}) and (\ref{eq:definitionU})] as a function of $I_{\mathrm{pr}}$ and for different values of the pump-pulse intensity $I_{\mathrm{pu}}$. The dependence of these phases on pulse intensities matches that exhibited by the time-delay-dependent features of $\mathcal{S}(\omega_{21},\tau)$ and $\mathcal{S}(\omega_{31},\tau)$ in Fig.~\ref{fig:gridInt} and in Figs.~\ref{fig:minimaphases}(c) and \ref{fig:minimaphases}(d) at positive time delays, confirming the validity of Eq.~(\ref{eq:tau-pupr}) for the interpretation of the phase of the oscillating features displayed by transient-absorption spectra. Also in this case, the minus sign in Eq.~(\ref{eq:tau-pupr}) results in a shift in the local-minimum points in Figs.~\ref{fig:minimaphases}(c) and \ref{fig:minimaphases}(d) in a direction which is opposite to the change in phase exhibited by Figs.~\ref{fig:minimaphases}(e) and \ref{fig:minimaphases}(f).

In contrast to the previously discussed probe-pump case, here, the first arriving pump pulse equally influences the shift in the spectra evaluated at $\omega_{21}$ and $\omega_{31}$. This could already be observed in Figs.~\ref{fig:gridfreq} and \ref{fig:gridInt}, and is now confirmed by Eq.~(\ref{eq:tau-pupr}). The same phase $\psi_{\mathrm{pu}}$ equally affects both spectra, with a common shift which quantifies the \textit{phase difference} between excited states generated by the first-arriving pump pulse. This is apparent by the definition of $\psi_{\mathrm{pu}}$ [Eq.~(\ref{eq:psipu})] and of the associated term $Z_{\mathrm{pu}} = U_{\mathrm{pu},21}U_{\mathrm{pu},31}^*$ [Eq.~(\ref{eq:ZAB})], which represents the coherence between excited states $|2\rangle$ and $|3\rangle$ resulting from the interaction with the pump pulse, as schematically illustrated in Fig.~\ref{fig:schematic-explanation}. 

Quantifying the shift in the spectra induced by the second-arriving probe pulse is more complex. In a pump-probe setup, the probe pulse modifies the state excited by the first-arriving pump pulse, inducing atomic phase changes which are encoded in the spectrum. However, in this case, the phases $\psi_{\mathrm{pr},k}$ [Eq.~(\ref{eq:psipr})] of the time-delay-dependent oscillations of $\mathcal{S}(\omega_{k1},\tau)$ are due to a sum of terms [$(A_{\mathrm{pr},k}-B_{\mathrm{pr},k}^*)$ from Eq.~(\ref{eq:ZAB})]. As a result, the phases $\psi_{\mathrm{pr},k}$, and hence the corresponding phase shifts featured by the spectra, are not only determined by the phases of the corresponding interaction-operator matrix elements ($U_{\mathrm{pr},12}U_{\mathrm{pr},k3}$ and $U_{\mathrm{pr},13}U_{\mathrm{pr},k2}$), but also by their amplitudes. The definition of the interaction operator $\hat{U}(I)$ allows one to see that $A_{\mathrm{pr},k}$ and $B_{\mathrm{pr},k}^*$ describe how the probe pulse transforms an initial coherence between the excited states $|2\rangle$ and $|3\rangle$ into a coherence between ground state and excited state $|k\rangle$ (see also the schematic illustration in Fig.~\ref{fig:schematic-explanation}). Amplitude and phase of these interaction-operator matrix elements both enter the definition of $\psi_{\mathrm{pr},k}$ and are hence encoded in intensity- and time-delay-dependent transient-absorption spectra.

\subsubsection{Dependence on laser frequency}

Since we confirmed in the previous subsections the validity of Eqs.~(\ref{eq:tau-prpu}) and (\ref{eq:tau-pupr}) for the interpretation of transient-absorption spectra in terms of pump- and probe-pulse-generated phases, here we focus on the previously introduced phases $\varphi_{\mathrm{pr},k}$, $\varphi_{\mathrm{pu},k}$, $\psi_{\mathrm{pu}}$, and $\psi_{\mathrm{pr},k}$, and investigate their dependence upon the frequency of the laser. Also in this case, this is achieved by using Eqs.~(\ref{eq:phi}), (\ref{eq:phiweak}), (\ref{eq:psipu}), (\ref{eq:psipr}), and (\ref{eq:psiweak}), after having numerically calculated $\hat{U}_{\mathrm{pr}}$ and $\hat{U}_{\mathrm{pu}}$ via Eqs.~(\ref{eq:matrixU0}) and (\ref{eq:definitionU}). However, while we assumed in the previous sections that both pump and probe pulses were characterized by a laser frequency $\omega_{\mathrm{L}} = 1.59\,\mathrm{eV}$, we display here intensity-dependent results for 5 discrete values of laser frequency, equally spaced between $\omega_{\mathrm{L}} = \omega_{21} = 1.56\,\mathrm{eV}$ and $\omega_{\mathrm{L}} = \omega_{31} = 1.59\,\mathrm{eV}$. 

\begin{figure}[t]
\includegraphics[width=\columnwidth]{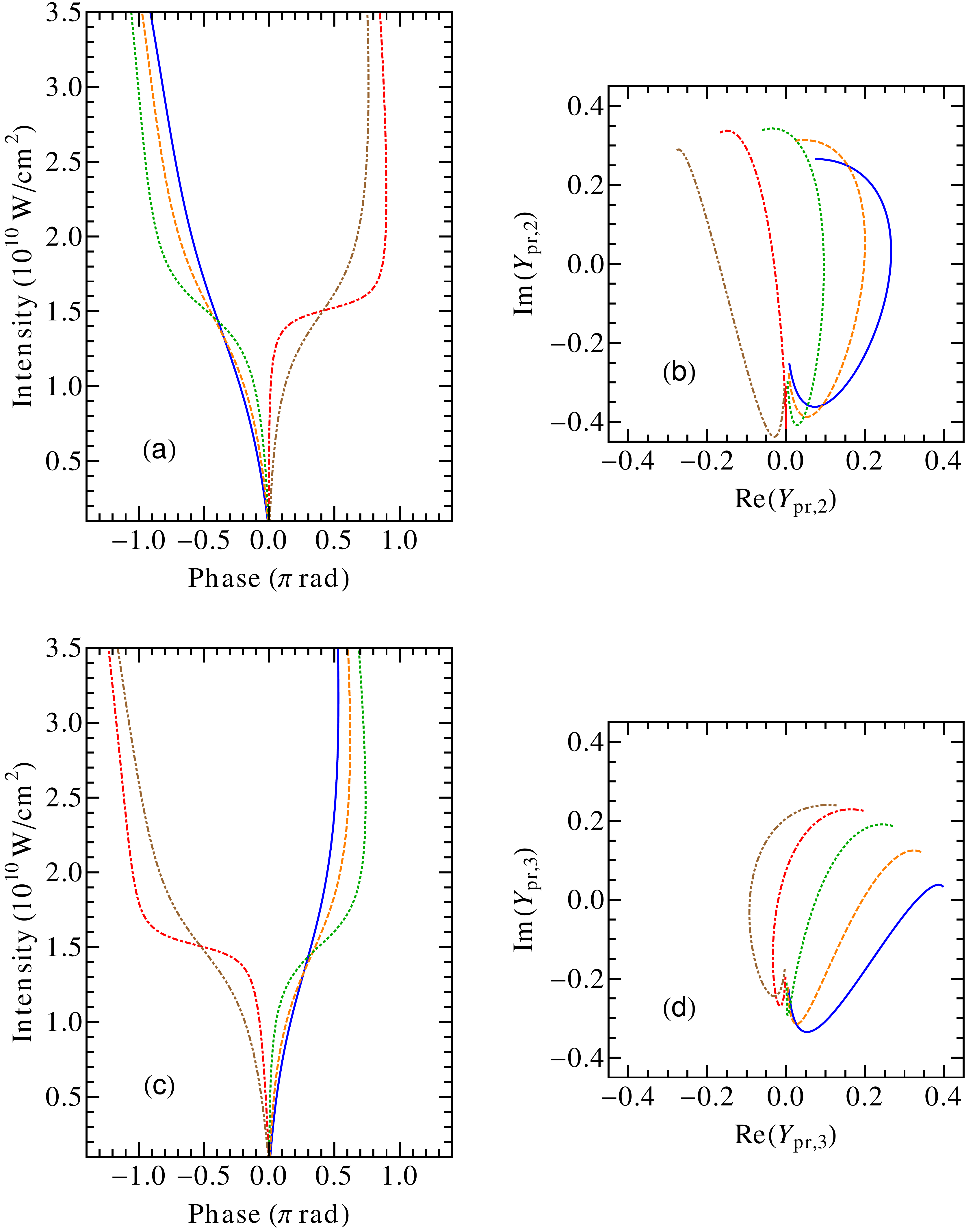}
\caption{Probe-pulse-induced phases in a probe-pump setup as a function of probe-pulse intensity and for
$\omega_\mathrm{L} = 1.560\,\mathrm{eV}$ (blue continuous line), 
$\omega_\mathrm{L} = 1.567\,\mathrm{eV}$ (orange dashed line), $\omega_\mathrm{L} = 1.575\,\mathrm{eV}$ (green dotted line), 
$\omega_\mathrm{L} = 1.582\,\mathrm{eV}$ (red dashed-dotted line), $\omega_\mathrm{L} = 1.590\,\mathrm{eV}$ (brown dashed-double-dotted line). The panels display (a) $(\varphi_{\mathrm{pr},2} -\varphi_{\mathrm{pr},2}^{\mathrm{weak}})$ and (c) $(\varphi_{\mathrm{pr},3} -\varphi_{\mathrm{pr},3}^{\mathrm{weak}})$, with the real and
imaginary parts of the corresponding complex numbers (b) $Y_{\mathrm{pr},2}$ and (d) $Y_{\mathrm{pr},3}$.
}
\label{fig:UprPRPU}
\end{figure}

\begin{figure}[t]
\includegraphics[width=\columnwidth]{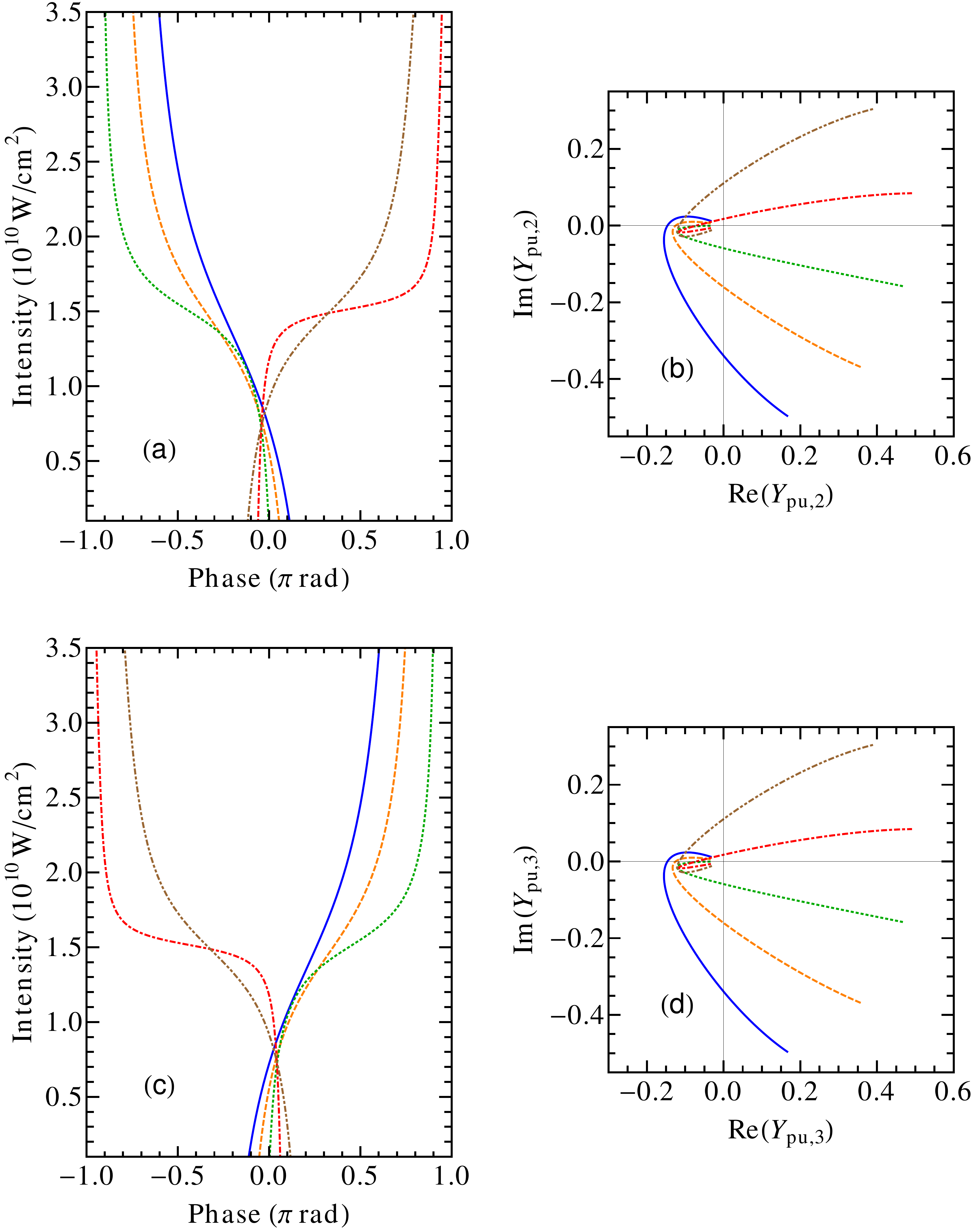}
\caption{Pump-pulse-induced phases in a probe-pump setup as a function of pump-pulse intensity and for
the same laser frequencies employed in Fig.~\ref{fig:UprPRPU}. The panels display (a) $(\varphi_{\mathrm{pu},2} -\varphi_{\mathrm{pu},2}^{\mathrm{weak}})$ and (c) $(\varphi_{\mathrm{pu},3} -\varphi_{\mathrm{pu},3}^{\mathrm{weak}})$, with the real and
imaginary parts of the corresponding complex numbers (b) $Y_{\mathrm{pu},2}$ and (d) $Y_{\mathrm{pu},3}$.
}
\label{fig:UpuPRPU}
\end{figure}

In Figs.~\ref{fig:UprPRPU} and \ref{fig:UpuPRPU}, we focus on a probe-pump setup and display the phases induced by the probe and pump pulses, respectively, as a function of their intensity and for different laser frequencies. Figure~\ref{fig:UprPRPU}(a) shows the phase $(\varphi_{\mathrm{pr},2} -\varphi_{\mathrm{pr},2}^{\mathrm{weak}})$ which determines the probe-intensity-dependent shift featured by the absorption spectra $\mathcal{S}(\omega_{21},\tau)$ evaluated at $\omega_{21}$. These phases are related to the argument of $Y_{\mathrm{pr},2}$ [Fig.~\ref{fig:UprPRPU}(b) and Eq.~(\ref{eq:Y})], which represents the coherence between states $|1\rangle$ and $|3\rangle$ generated by the first-arriving probe pulse. At low intensities, all curves are characterized by negative, purely imaginary values of $Y_{\mathrm{pr},2}$, in agreement with Eq.~(\ref{eq:Yweak}). The laser frequency influences the path followed by $Y_{\mathrm{pr},2}$ at increasing intensities, and whether this will move towards regions characterized by positive or negative real parts. This influences the behavior of the phases in Fig.~\ref{fig:UprPRPU}(a) as well, deciding whether the shift is towards values of $\varphi_{\mathrm{pr},2}$ larger or smaller than the weak-limit value. Similarly, the behavior of $Y_{\mathrm{pr},3}$ displayed in Fig.~\ref{fig:UprPRPU}(d) determines the intensity-dependent shift $(\varphi_{\mathrm{pr},3} -\varphi_{\mathrm{pr},3}^{\mathrm{weak}})$ featured by the absorption spectra $\mathcal{S}(\omega_{31},\tau)$ evaluated at $\omega_{31}$. Here, $Y_{\mathrm{pr},3}$ is the coherence between states $|1\rangle$ and $|3\rangle$ generated by the first-arriving probe pulse. Also in this case, weak intensities correspond to negative, purely imaginary values of $Y_{\mathrm{pr},3}$, agreeing with Eq.~(\ref{eq:Y}). A different dependence of $Y_{\mathrm{pr},3}$ on probe-pulse intensities is featured for different values of the laser frequency, analogously influencing the intensity-dependent shift $\varphi_{\mathrm{pr}, 3}$ exhibited by Fig.~\ref{fig:UprPRPU}(c).

Figure~\ref{fig:UpuPRPU} shows the additional phase shift owing to a strong pump pulse as a function of its intensity. The (amplitude and phase) changes resulting from the interaction with the pump pulse are encoded in the complex numbers $Y_{\mathrm{pu},2}$ and $Y_{\mathrm{pu},3}$, whose dependence on intensity and laser frequency is shown in Figs.~\ref{fig:UpuPRPU}(b) and \ref{fig:UpuPRPU}(d), respectively. As noticed in Eq.~(\ref{eq:Yweak}), $Y_{\mathrm{pu},k}$ are of second order in the pulse area $\vartheta$ for weak values of the pulse intensity. As a result, for small pulse intensities, $Y_{\mathrm{pu},2}$ and $Y_{\mathrm{pu},3}$ tend to vanishing values for all considered laser frequencies. The associated atomic-phase change results in the phase shifts displayed in Figs.~\ref{fig:UpuPRPU}(a) and \ref{fig:UpuPRPU}(c). For all considered laser frequencies, $\varphi_{\mathrm{pu},2}$ and $\varphi_{\mathrm{pu},3}$ evolve in opposite directions for increasing values of the pump-pulse intensity.

\begin{figure}[t]
\includegraphics[width= \columnwidth]{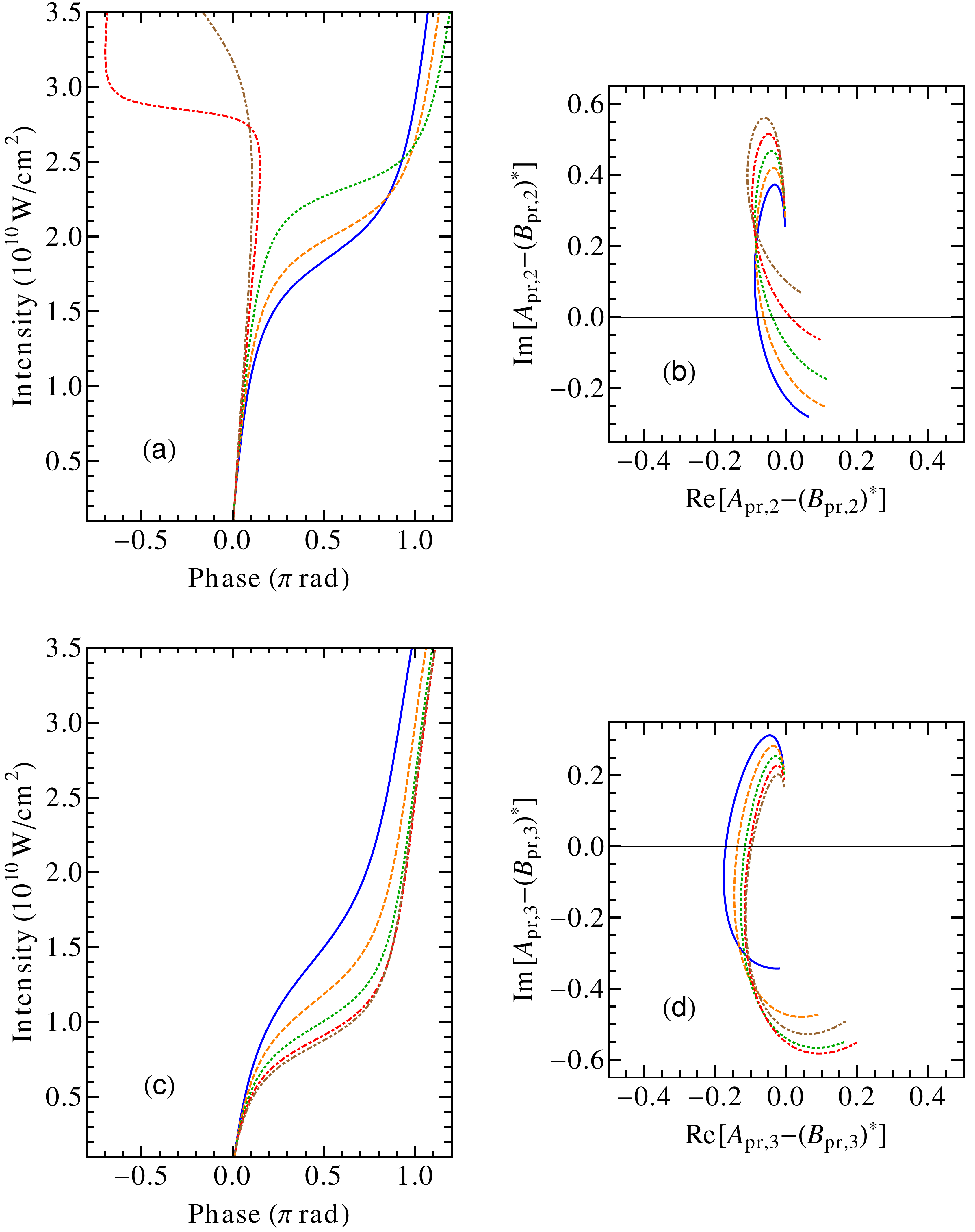}
\caption{Probe-pulse-induced phases in a pump-probe setup as a function of probe-pulse intensity and for
the same laser frequencies employed in Fig.~\ref{fig:UprPRPU}. The panels display (a) $(\psi_{\mathrm{pr},2} - \psi_{\mathrm{pr},2}^{\mathrm{weak}})$ and (c) $(\psi_{\mathrm{pr},2} - \psi_{\mathrm{pr},2}^{\mathrm{weak}})$, with the real and
imaginary parts of the corresponding complex numbers (b) $(A_{\mathrm{pr},2} - B_{\mathrm{pr},2}^*)$ and (d) $(A_{\mathrm{pr},3} - B_{\mathrm{pr},3}^*)$.
}
\label{fig:UprPUPR}
\end{figure}

\begin{figure}[t]
\includegraphics[width= \columnwidth]{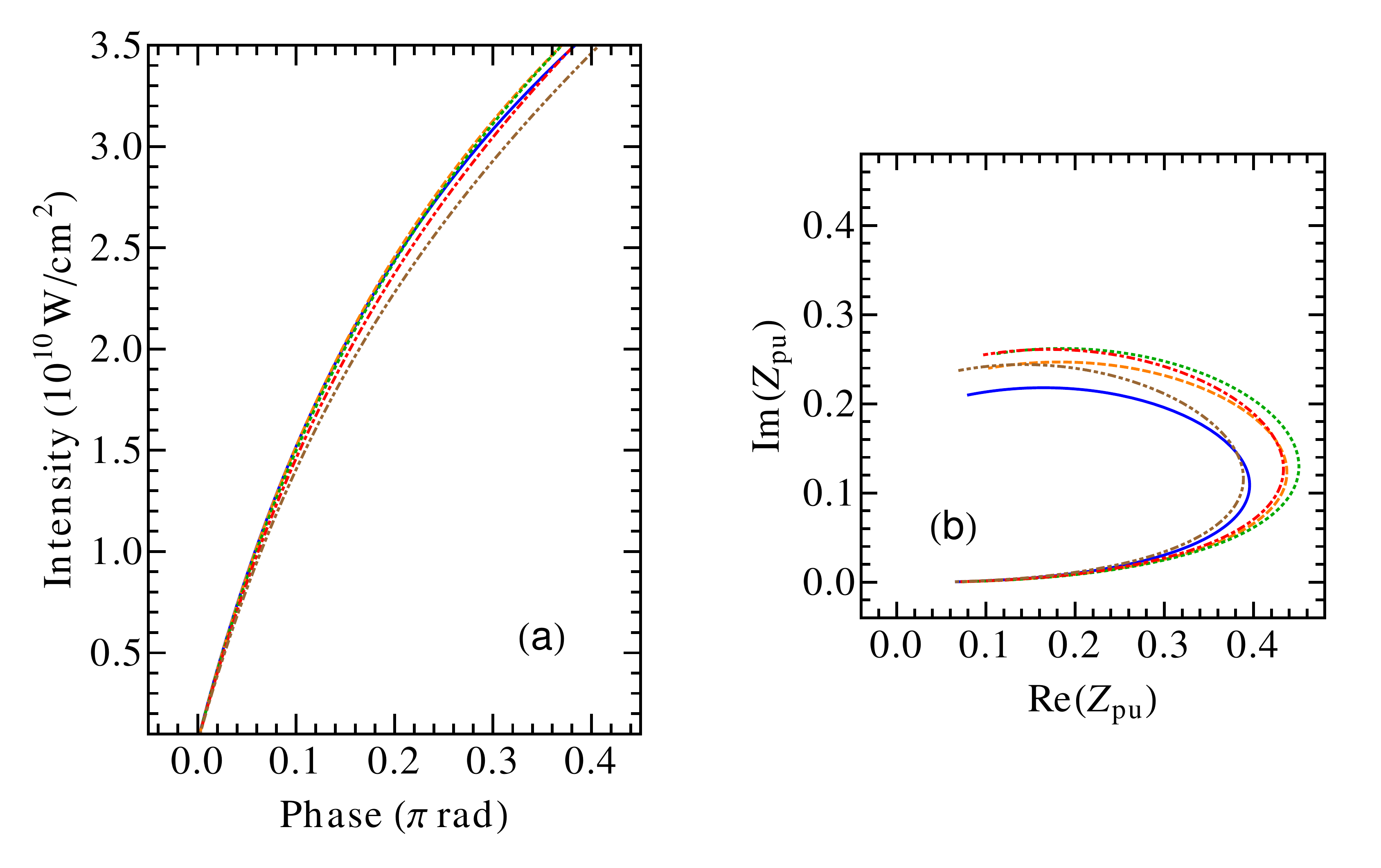}
\caption{Pump-pulse-induced phases in a pump-probe setup as a function of pump-pulse intensity and for
the same laser frequencies employed in Fig.~\ref{fig:UprPRPU}. The panels display (a) $(\psi_{\mathrm{pu}} - \psi_{\mathrm{pu}}^{\mathrm{weak}})$ and (b) the real and
imaginary parts of the corresponding complex number $Z_{\mathrm{pu}}$.
}
\label{fig:UpuPUPR}
\end{figure}

In Figs.~\ref{fig:UprPUPR} and \ref{fig:UpuPUPR} we consider a pump-probe setup and focus on the phases induced by probe and pump pulses, respectively, as a function of their intensity and for different laser frequencies. The phase $(\psi_{\mathrm{pr},2} -\psi_{\mathrm{pr},2}^{\mathrm{weak}})$, defining the intensity-dependent shift of $\mathcal{S}(\omega_{21},\tau)$, is shown in Fig.~\ref{fig:UprPUPR}(a). Also in this case, the displayed dependence upon intensity and laser frequency can be better understood by referring to the complex numbers $(A_{\mathrm{pr},2} - B^*_{\mathrm{pr},2})$ [Eq.~(\ref{eq:ZAB})], displayed in Fig.~\ref{fig:UprPUPR}(b). As discussed previously, these complex numbers are related to the transformation induced by the second arriving-probe pulse, quantifying how an initial coherence between states $|2\rangle$ and $|3\rangle$ is transformed into coherence between $|1\rangle$ and $|2\rangle$. At low intensities, all curves tend to positive, purely imaginary values of $(A_{\mathrm{pr},2} - B^*_{\mathrm{pr},2})$, in agreement with Eq.~(\ref{eq:ZABweak}). The path followed by $(A_{\mathrm{pr},2} - B^*_{\mathrm{pr},2})$ at increasing intensities depends on the laser frequency, and reveals interesting features about the intensity dependence of $(\psi_{\mathrm{pr},2} -\psi_{\mathrm{pr},2}^{\mathrm{weak}})$ shown in Fig.~\ref{fig:UprPUPR}(a). For example, one can notice how relatively similar values of $(A_{\mathrm{pr},2} - B^*_{\mathrm{pr},2})$, such as those displayed by the green, red, and brown curves in Fig.~\ref{fig:UprPUPR}(b), can lead to a very different behavior of the corresponding phases [Fig.~\ref{fig:UprPUPR}(a)]. This is due to the fact that the amplitude of $(A_{\mathrm{pr},2} - B^*_{\mathrm{pr},2})$ is very close to vanish for all three considered curves. A small change in the actually followed path can therefore lead to a completely oppositely directed shift in the corresponding phase. The phases $(\psi_{\mathrm{pr},3} -\psi_{\mathrm{pr},3}^{\mathrm{weak}})$ shown in Fig.~\ref{fig:UprPUPR}(c), determining the intensity-dependent shift of $\mathcal{S}(\omega_{31},\tau)$, display a more regular dependence upon intensity and laser frequency. This is essentially related to the fact that the corresponding complex numbers $(A_{\mathrm{pr},3} - B^*_{\mathrm{pr},3})$ do not approach vanishing values for the range of intensities and laser frequencies considered, as exhibited by Fig.~\ref{fig:UprPUPR}(d). The complex numbers $(A_{\mathrm{pr},3} - B^*_{\mathrm{pr},3})$ [Eq.~(\ref{eq:ZAB})] quantify how an initial coherence between states $|2\rangle$ and $|3\rangle$ is transformed into coherence between $|1\rangle$ and $|3\rangle$ and, at low intensities, tend to positive, purely imaginary values [Fig.~\ref{fig:UprPUPR}(d)], in agreement with Eq.~(\ref{eq:ZABweak}).

Finally, Fig.~\ref{fig:UpuPUPR} shows the pump-pulse-induced phase shift $(\psi_{\mathrm{pu}} - \psi_{\mathrm{pu}}^{\mathrm{weak}})$ which equally affects the oscillations of $\mathcal{S}(\omega_{21},\tau)$ and $\mathcal{S}(\omega_{31},\tau)$, as described in Eq.~(\ref{eq:tau-pupr}) for positive time delays. The associated complex numbers $Z_{\mathrm{pu}}$, quantifying the coherence between excited states generated by the first-arriving pump pulse, are exhibited in Fig.~\ref{fig:UpuPUPR}(b), displaying a small dependence on the laser frequency $\omega_{\mathrm{L}}$. This is reflected in the associated phases, shown in Fig.~\ref{fig:UpuPUPR}(a). We notice that $Z_{\mathrm{pu}}$ tends to 0 for small intensities, being of second order in the pulse area $\vartheta$ as predicted by Eq.~(\ref{eq:ZABweak}). For increasing values of the intensity, however, we can see that $Z_{\mathrm{pu}}$ is characterized by values very close to the real axis, in agreement with the prediction of a vanishing weak-limit phase $\psi_{\mathrm{pu}}^{\mathrm{weak}} = 0$ [Eq.~(\ref{eq:psiweak})].

\section{Conclusion}
\label{Conclusion}
In conclusion, we have investigated the interaction of a sample of Rb atoms, modeled as a $V$-type three-level system,
with intense probe and pump pulses separated by a positive or negative time delay in a transient-absorption-spectroscopy setup.
The three-level model was used to describe the evolution of the atomic system and, thereby, to
numerically simulate experimental time-delay- and pulse-intensity-dependent
spectra.
We developed an analytical interpretation model, which we used to connect the time-delay-dependent oscillations featured by the spectra with the pump- and probe-pulse-induced quantum phases of the atomic system. Thereby, we showed which strong-field information on atomic phases can be extracted from transient-absorption spectra, when intense probe and pump pulses are employed. 
We also studied the dependence of strong-field-generated atomic phases on the frequency of the utilized laser pulses.

Further studies could include a more thorough analytical and theoretical description of the frequency
dependence of the phases, as well as an atomic-system description going beyond the three-level model employed here. For high densities or long media, it could be important to further investigate how propagation effects can be included in our interpretation models. 

\begin{acknowledgments}
The authors acknowledge valuable discussions with Zolt\'an~Harman, Christoph~H.~Keitel, and Thomas~Pfeifer. The work of V.~B. has been carried out thanks to the support of the A*MIDEX grant (No.~ANR-11-IDEX-0001-02) funded by the French Government <<Investissements d'Avenir>> program.
\end{acknowledgments}

\end{document}